\documentstyle[12pt,aj_pt4]{article}

\def\etal{{\it et al.\,}}

\def\logg{log {\it g}}
\def\vrf{${\rm v}_{\rm RF}$}
\def\teff{${\rm T}_{\rm eff}$}

\received{December 14, 2000}
\accepted{September 28, 2001}
\slugcomment{Accepted for publication in January 2002 AJ}

\lefthead{Fulbright}
\righthead{Abundances \& Kinematics II}

\begin{document}{
\title{Abundances and Kinematics of Field Stars II:\\ Kinematics and Abundance Relationships}

\author{Jon P. Fulbright\altaffilmark{1}} 

\affil{Dominion Astrophysical Observatory, Herzberg Institute of Astrophysics, National Research Council, 5071 West Saanich Road, Victoria, BC V8X-4M6, Canada.  Email:  Jon.Fulbright@nrc.ca}

%\affil{UCO/Lick Observatory, Dept of Astronomy, University of California, Santa Cruz, CA 95064}

\begin{abstract}
As an investigation of the origin of ``$\alpha$-poor'' halo stars, we 
analyze kinematic and abundance data for 73 intermediate metallicity stars
($-1 >$ [Fe/H] $\geq -2$) selected from Paper I of this series.  We find 
evidence for a connection between the kinematics and the enhancement
of certain element-to-iron ([X/Fe]) ratios in these stars.  Statistically 
significant correlations were found between [X/Fe] and galactic rest-frame 
velocities (\vrf{}) for Na, Mg, Al, Si, Ca and Ni, with marginally 
significant correlations existing for Ti and Y as well.  We also find that 
the [X/Fe] ratios for these elements all correlate with a similar level of
significance with [Na/Fe].  Finally, we compare the abundances of these halo
stars against those of stars in nearby dSph galaxies.  We find significant 
differences between the abundance ratios in the dSph stars and halo stars of 
similar metallicity.  From this result, it is unlikely that the halo stars in 
the solar neighborhood, including even the ``$\alpha$-poor'' stars, were once 
members of disrupted dSph galaxies similar to those studied to date.
\end{abstract}


\section{Introduction}

Traditionally, the chemical enrichment history of the Galaxy is described
to follow the general scenario given by \cite{t97}.  In this scenario, the
metal-poor halo stars formed from the ejecta of short-lived Type II supernovae.
These events eject material rich in the so-called ``$\alpha$-elements'' (here
taken as O, Mg, Si, Ca, and Ti).  Observed halo stars, which presumably formed
from this gas, preserved this pattern by showing [$\alpha$/Fe] ratios of 
$\sim +0.3$.  About 1 Gyr after the initial burst, Type Ia supernovae began 
exploding, ejecting Fe-group rich material.  This explains the observed drop 
of the [$\alpha$/Fe] ratio from the halo value at [Fe/H] $\sim -1$ to solar at 
[Fe/H] $\sim 0$ (Edvardsson \etal 1993; \cite{m97}).

More recently studies have found halo stars that do not follow
this pattern of $\alpha$-element enhancement.  \cite{c97}, \cite{k97}, and 
\cite{h98} all found stars with [$\alpha$/Fe] ratios lower than what is 
normally expected for halo stars.  The globular clusters Ruprecht 106 and 
Palomar 12 (\cite{b97}) also show [$\alpha$/Fe] ratios that are relatively low
(for their metallicity) compared to the majority of the halo.  The most 
exceptional case is BD +80 245, which shows near-solar [$\alpha$/Fe] ratios at 
[Fe/H] $\sim -2$ (Carney et al. 1997).  These ratios suggest a different 
chemical enrichment history for these stars than for the rest of the halo.  

Kinematics may help determine the origin of these stars.  
\cite{c97} noted that the known metal-poor halo stars with 
unusual abundances all have high apogalactic radii, while H98 
found that more of the stars with lower [Na/Fe] ratios move on retrograde 
orbits.  It should be possible, using the large,       
self-consistent sample of \cite{f00} to test whether or not stars on
extreme orbits demonstrate unusual element abundance ratios.  

The F00 data set includes the LTE analysis of 15 elements based on
high resolution, high signal-to-noise echelle spectra of 168 stars.
These stars were selected primarily by metallicity from lists of
known metal-poor stars or by their potential as being subdwarfs from
their location on a Hipparcos-based color-magnitude diagram.  All of the 
observed stars are members of the Hipparcos catalog.  
In addition to these 168 stars, an additional 11 stars with equivalent width 
measurements from \cite{s99}, were reanalyzed following the same procedure.
The analysis procedure of F00 followed a self-consistent methodology aimed at 
producing the most accurate abundance measurements possible.

One suggestion for the origin of the $\alpha$-poor stars is that these stars 
may have been accreted since the formation of the rest of the stellar halo.  
\cite{shw92} and \cite{gw98} have suggested that dwarf galaxies or 
proto-galactic fragments may have chemical enrichment histories with 
lower $\alpha$-element enhancements.  While it not possible to determine
absolutely the origin of individual field stars, the abundances of stars 
within potential future accretion targets (e.g., Milky Way satellite 
dSph galaxies) should follow patterns similar to what Smecker-Hane 
\& Wyse or Gilmore \& Wyse suggest.  The recent abundance analysis
of \cite{s01} of stars within 3 nearby dSph galaxies allows such comparisons 
to be made in one potential source of accreted stars.

This paper will be organized as follows:
In Section 2 of this paper, we will calculate kinematic parameters for 
the F00 stars and select a sample for further study.  The data will 
then be used to demonstrate that there exist correlations between 
abundance ratios and kinematics and that there are correlations between 
the element enhancements of certain light elements (Section 3).  
Finally, in Section 4 we will show that true analogues to the 
$\alpha$-poor halo stars do not exist in the Milky Way satellite 
dSph galaxies studied to date.

\section{Data Used in the Analysis}

\subsection{Abundance Data}

The abundance ratios from F00 are presented in Figures 1--3.  Also plotted
are stars from several recent abundance surveys.  As can be seen, there is
significant overlap for metal-rich stars between the F00 data and the data 
of \cite{edv93} and \cite{cl99}.  In the extremely 
metal-poor regime ([Fe/H] $\lesssim -2$), the \cite{rnb} and the 
\cite{mcw95} data provide more stars than F00, but the intermediate
metallicity range ($-2 <$ [Fe/H] $< -1$) is nearly exclusively covered by 
the F00 data.  
	
Figure 4 presents comparisons of a few of the abundance ratios measured by
both F00 and the other surveys.  There is an overall good agreement between 
the F00 results and previous surveys, except for a $\sim 0.1$ dex offset 
in [Si/Fe].  The origin of this offset is unclear--the atomic data used for
Si by the different surveys all agree well, and the sensitivity of [Si/Fe]
to systematic errors in the stellar parameters (see Table 8 of F00) is
smaller than the values measured for other [X/Fe] ratios that do not show
a similar offset.

\subsection{Kinematic Data}
 
	As stated in F00, one of the selection criteria for this survey is 
inclusion in the Hipparcos Catalog (\cite{hip}).  All of the proper motions 
and positions used in this survey come from the catalog.  With the
exception of several giants whose Hipparcos parallaxes are uncertain,
all of the adopted distances are derived from Hipparcos parallaxes.  We adopt 
the H98 or \cite{att} distances (giving preference to H98) for those giants 
with Hipparcos $\delta(\pi)/\pi > 0.2$.  

        The adopted radial velocities (see Table 1) come from several sources, 
most often the \cite{clla} survey and the Hipparcos Input Catalog (\cite{hic}).
Other radial velocities were adopted from the literature.  For the remaining 
stars we measured radial velocities from the spectra themselves.  No 
calibrations to radial velocity standards were made, but for the 73 stars
in common with \cite{clla}, the mean difference in the observed heliocentric 
radial velocity (F00 $-$ CLLA) is $+0.3 \pm 3.9$ km $s^{-1}$ (sdom), and for the 38 
stars with radial velocities adopted from the Hipparcos Input Catalog the 
mean difference (F00 $-$ HIC) is $+1.9 \pm 4.4$ km $s^{-1}$.  

        Given the above input parameters, the UVW velocities for the stars
were calculated using a set of programs kindly provided by R. Hanson 
(private communication).  The UVW velocities are defined such that positive
U values denote motion away from the Galactic Center, positive V values are
in the direction of the solar motion, and positive W values are parallel to 
the direction of the North Galactic Pole.  The UVW components were 
transformed to the local standard of rest (LSR) using the solar motion 
$(U_{\odot}$,$V_{\odot}$,$W_{\odot})_{LSR} = (-9,12,7$ km $s^{-1})$ 
(Mihalas and Binney, p. 400).  For purposes of this paper, the rotational 
velocity of the LSR with respect to the Galaxy was set to 220 km/$s^{-1}$ and
the galactocentric distance was set to 8.5 kpc.  
Also given in Table 1 are the galactic rest frame velocities, 
\vrf{} ($= \sqrt{(U_{LSR}^2 + (V_{LSR} + 220)^2 + W_{LSR}^2)}\;$), orbital
velocities in the direction of the Sun's orbit, $v_{ROT}$ 
($= V_{LSR}+220$ km $s^{-1}$) and specific angular momentum values, 
$h$ ($= \sqrt{(v_{ROT}^2 + W_{LSR}^2)}\;$).   

        Orbital parameters for each star (maximum and minimum galactocentric
radii, $R_{max}$ and $R_{min}$, maximum absolute distance from the galactic 
plane, $|Z_{max}|$,  and orbital eccentricity, $e$) were calculated using an 
integrator kindly provided by 
D. Lin (1999, private communication).  The integrator uses a three component 
potential describing the halo, disk, and bulge and is described in more 
detail by \cite{j98}.  Each star was followed for 5 Gyr or at least 8 orbits.  
Table 1 lists the resulting kinematic and orbital parameter values.  For 
completeness, the stars from S99 are listed with kinematic parameters 
from the \cite{clla} survey, but the orbital parameters were recalculated using
the same galactic potential model as the F00 stars.

\subsection{Selection of Stars for Further Study}

An important kinematic parameter is the rest-frame velocity of
the star with respect to the center of mass of the Galaxy (\vrf{}).  
\cite{c99} notes that the known $\alpha$-poor stars show high galactic 
rest-frame velocities.  If we wish to test whether this is true, we 
need to ensure that our sample does not exclude these important stars.  
In Figure 5 we plot [Fe/H] against \vrf{} and find that the F00 data 
set does not include a large number of very metal-poor ([Fe/H] $< -2$), 
high-velocity stars.  This is most likely a selection effect, as the 
high-velocity star study of \cite{c88} found that out of a sample of 24 
stars with \vrf $> 375$ km/$s^{-1}$, 13 have $-1 >$ [Fe/H] $\geq -2$ and 10 
showed [Fe/H] $< -2$.  Most of the very metal-poor, high-velocity stars
included in \cite{c88} were too faint to be included in the F00 survey,
and only the availability of the S99 data made it possible to include a 
reasonable number of high-velocity stars.

Also included in Figure 5 are stars from \cite{edv93}
sample of disk stars.  As can be seen, for the metal-rich stars 
([Fe/H] $> -1$) the F00 and \cite{edv93} samples share the same regions.  
Similar comparisons of the other kinematic and orbital parameters between the 
two samples in this metallicity range lead us to believe that the stars with 
[Fe/H] $> -1$ in the F00 sample are almost all members of the disk population.

Therefore, to ensure a less biased comparison of stars, we will generally 
confine our discussion of the relationship between kinematics and abundances 
to the 73 stars with $-1 <$ [Fe/H] $< -2$, as demarked by the dotted lines 
in Figure 5.  For simplicity, we will refer to this set of stars as 
the ``selected sample'' for the remainder of the paper.  Note that the
selected sample does not include the well-know $\alpha-poor$ star BD $+80 245$.

\section{Abundances as a Function of Kinematic Properties}

\subsection{[X/Fe] vs. \vrf{}}

In this section, we will test whether stars showing extreme orbital parameters
show different abundance parameters than stars on more normal orbits.  A
useful kinematic parameter for these comparisons is \vrf{}, as defined above.
This parameter is directly related to the star's kinetic energy with respect
to the Galaxy.  The value of \vrf{} is determined solely from observational
data and, unlike $R_{max}$, is independent of the model is used to
describe the galactic gravitational potential.

As a quantitative way to describe any trends with \vrf{}, we fit least-squares
lines to the distributions of [X/Fe] vs. \vrf{} for the selected sample.  In
Figures 6 and 7 we plot the fits to the eight element ratios that are somewhat
significant.  For the fits involving Na, Mg, Al, Si, Ca, and Ni, the value of
the correlation coefficients are such that there is a less than 0.05\% chance 
that the correlations are random.  For the fits involving Ti and Y, the 
probability is a few percent.  These  significance levels were confirmed by a 
non-parametric Spearman rank-order 
analysis, which compares only the rank order of the two variables and is 
independent of the form of the correlation.  Fits to the other element ratios 
(including [Fe/H]) did not show any significant correlations.  For simplicity,
we will refer to the [X/Fe] ratios for Na, Mg, Al, Si, Ca and Ni as the
``varying'' ratios. 

To further explore the potential relationship between [X/Fe] and \vrf{},
we divide the selected sample into three \vrf{} groups.  The properties 
of these three groups are given in Table 2 and the mean values of [X/Fe] 
are plotted in Figure 8.  The three groups have a similar mean \teff{} 
and [Fe/H] values, but the highest velocity group has a higher mean value
of \logg{}.  This is due to the inclusion of the S99 stars to help
fill out the highest velocity group.  Even among metal-poor stars high 
velocity stars are rare.  The S99 stars were specifically selected for
their kinematics from the \cite{clla} survey, which primarily included
dwarf stars.  The F00 survey selection was mainly based on metallicity 
which allowed more evolved stars into the survey.  

Figure 8 demonstrates that the highest velocity stars have a different
distribution of light elements.  The mean values of both [Na/Fe] and [Mg/Fe] 
are $\sim 0.2$ dex lower for the high velocity group, while the mean
values of [Al/Fe], [Si/Fe], [Ca/Fe], [Y/Fe] and [Ba/Fe] are slightly lower 
as well.  The error bars in Figure 8 represent the standard deviation of the 
mean within each bin for that element ratio.  They do not include the estimated 
random errors in the observations of the ratios ($\sim 0.1$ dex, as 
discussed in F00).

Besides the relatively low light element abundance ratios, it is possible that
the highest velocity stars also show slightly lower ratios of the s-process
elements Y, Zr, and Ba.  The mean [Ba/Eu] ratio does show a change between
the velocity groups, but this measurement is limited by the number of Eu
measurements.  A lower s-process fraction in these stars would suggests that 
chemical evolution history of these stars contains less recycling of the 
ejecta from AGB stars, the suspected site of the main s-process.  The size
of the s-process deficiency is small and may not be significant, so an 
enlarged sample of high-quality s-process element ratios should help clarify 
this picture.

\subsection{Element-Element Correlations}

        The trends in Figures 6--8 suggest that the light element ratios
may be correlated with each other.  If this is the case, then [Na/Fe], which
shows the largest range star-to-star variations in [X/Fe], could be used as 
a surrogate for describing the overall enhancements of these element
ratios.  Indeed, H98 and \cite{s98} present evidence that there is a 
correlation between [Na/Fe] and [Mg/Fe] in halo giant stars.  Here we attempt 
to expand this to other element ratios.  

For the selected sample, we made linear least-squares fits to the [X/Fe] vs.
[Na/Fe] distributions.  As a comparison, fits were also made to the [X/Fe] 
vs. [Fe/H] distributions as well.  We find strongly significant relationships 
between [Na/Fe] and the other varying ratios.  Correlations 
significant to the few percent level were found for the fits with [Ti/Fe] and 
[Y/Fe].  For the fits using [Fe/H] as the independent variable, only [Cr/Fe] 
shows a strongly significant correlation.  The above significance levels were 
confirmed through the use of Spearman rank-order tests.  

Another confirmation comes from the relative size of the standard deviations
of the [X/Fe] values around the least squares fits ($\sigma_{\rm [X/Fe]}$).  
For the five ratios that strongly correlate with [Na/Fe], the values of 
$\sigma_{\rm [X/Fe]}$ found for the fit to [Na/Fe] are $\sim 30\%$
smaller than those found for the fit to [Fe/H].  For the remaining ratios, the
$\sigma_{\rm [X/Fe]}$ values were comparable between the fits to the different
independent variables.  

In Figure 9 we plot the [X/Fe] vs. [Na/Fe] distributions for the four
$\alpha$-elements studied in F00.  All four ratios show trends with
[Na/Fe], but as a further test, we calculated the value of [$\alpha$/Fe]
(the mean of the [X/Fe] ratios for Mg, Si, Ca and Ti) for the 65 stars 
in the selected sample having abundance measurements for all four elements. 
The plot of [$\alpha$/Fe] vs. [Na/Fe] is presented in the top panel of 
Figure 10.  The quality of the fit and the statistical significance is
better than the fits to the individual ratios themselves.  Also plotted
in Figure 10 are the [X/Fe] vs. [Na/Fe] distributions and fits for Al,
Ni and Y.

Random and systematic errors in the stellar parameters (\teff, \logg, etc.) 
could possibly lead to the correlations seen here.  Table 8 of F00 gives the
effects of specific changes in the stellar parameters on the final abundance
ratios.  For most of the varying elements, the sign of the changes all match,
meaning that the errors are correlated.  However, the magnitude of the changes
are too small with respect to the variations seen in the elemets.  For example,
a change off 150 K in \teff, a 0.2 dex change in \logg, or a 0.3 km $s^{-1}$ 
change in the microturbulent velocity results in a change of $\sim 0.05$ dex
or less in the resulting abundances.  The change in parameter values needed
to explain the range of [X/Fe] values seen in unreasonable.  Also, the line
used in F00 were specifically chosen to ensure that the individual lines of
a given element have consisten results as compared to the other lines of
that element over a wide range of stellar papameters.  Because of the care
taken to form the line list it is unlikely, short of problems with the
basic assumptions of the LTE, plane-parallel analysis, that the results seen
here are artifacts of line selection or the abundnace analysis.

\subsection{Abundances and Orbital Parameters}

Instead of using \vrf{} as the kinematic parameter, \cite{c97} and \cite{c99}
noted potential trends between $R_{max}$ or $h$ and abundances.  Using [Na/Fe]
as a surroagate for the other varying element ratios, we can test these 
trends with the F00 data set. Plots of [Na/Fe] against orbital parameters are 
shown in Figures 10(a)--(c).  The fraction of 
``Na-poor'' stars (those with [Na/Fe] $< -0.36$) increases for $R_{max} > 20$ 
kpc (9 of 22 stars) and $|Z_{max}| > 5$ kpc (5 of 10 stars).  From these 
counts, about a half of extreme halo stars should show lower element ratios.  
Conversely, 55\% (6 of 11) of stars with [Na/Fe] $< -0.36$ have 
$R_{max} > 40$ kpc, while only 6\% of the remaining sample has values of 
$R_{max}$ this high. Ten of the 11 Na-poor stars have \vrf{} $> 300$ km $s^{-1}$, 
while only 11 of the 113 other stars have this high a value of \vrf{}.   

        \cite{c99} suggested that the previously observed alpha-poor
stars also had high ``specific angular momentum'' ($h$) values.  For any
star, the maximum value of $|h|$ is \vrf{}.  For the 11 stars with [Na/Fe]
$< -0.36$, the mean value of $|h|$/\vrf{} is $0.78 \pm 0.07$ (sdom).  If the 
one star (G197-30) with an $|h|$/\vrf{} value of 0.17 is eliminated, the mean 
value increases to $0.84 \pm 0.04$.  This compares to a mean value of 
$0.66 \pm 0.04$ for the other 61 stars in the selected sample with [Na/Fe] 
abundances.

\section{Accretion Origin of the $\alpha$-poor Stars?}

So far we have shown that the highest velocity stars in the solar
neighborhood show a different pattern of abundance ratios than
the lower velocity stars.  This suggests that these stars did not 
form in the same manner than the rest of the halo.  One theory for 
the origin of the low-$\alpha$ halo stars is that they were accreted 
from the cannibalization of other stellar systems.  If the stellar 
halo of the Milky Way was made up of disrupted and accreted proto-galactic 
fragments, as suggested by \cite{ct00}, then studies of stars within 
present-day dwarf galaxies (assuming they are products of similar fragments) 
may help explain the abundance patterns seen in the halo.   \cite{s01} 
analyzed high-resolution spectra of 17 giants stars in the Draco, Ursa Minor 
and Sextans dwarf spheroidal (dSph) galaxies.  Their methodology is similar 
to the LTE analysis technique employed by F00, although the spectra were of 
lower S/N ($\sim$ 15--35) than F00.

In Figure 12, we plot the mean [X/Fe] values for the 10 \cite{s01} dSph 
stars with $-1 >$ [Fe/H] $\geq -2$ along with the 3 kinematic groups of the 
selected sample stars.  For several elements the mean values of the [X/Fe] 
ratios differ between the dSph stars and the halo stars.  It is also 
clear that the dSph abundance pattern does not match what is seen in 
$\alpha$-poor stars.  For example, the most extreme $\alpha$-poor star 
known, BD $+80 245$ (= HIP 40068) shows [Ba/Fe] and [Eu/Fe] ratios $\sim 1.5$ 
dex below halo stars of similar metallicity (see Figure 3).  This would place 
it off the bottom of Figure 12, while the mean dSph values for these two ratios 
is higher than the mean halo value.

Since the F00 
survey did not find any stars with abundance pattern similar to the dSph 
stars, it is unlikely that a large fraction of the field halo stars in the
solar neighborhood, including the $\alpha$-poor stars, are former members 
of dSph galaxies of the kind studied by \cite{s01}.  

Despite this, continued high-resolution work on individual stars in other 
galaxies may find an extragalactic origin for the $\alpha$-poor stars.
For example, \cite{h00} present O and Al abundances derived from 
high-quality high-resolution spectra of 10 giants in globular clusters 
associated with the LMC.  For the stars that do not show signs of 
deep mixing (enhanced [Al/Fe] and depressed [O/Fe]), the general trend of
the metal-poor stars is to have low values of [O/Fe] and [Al/Fe] compared
to unmixed Galactic globular cluster giants.  The analysis of a full range
of elements in the stars is necessary before further comparisons can be made.
However, one exciting prospect is that the LMC cluster system shows a range of
ages (see \cite{ef88} and \cite{g97}), so it may be possible to explore the 
effects of time of formation on abundance ratios.

\section{Discussion and Future Work}

In this paper, we have presented observational evidence that shows that the
[X/Fe] ratios for certain elements decline in concert for higher values
of \vrf{}.  Here we discuss some possible scenarios that may produce
the pattern of elements found in the high velocity stars.  

        The production of Na and Mg by Type II supernova is dominated by 
massive progenitor stars with M $> 30 M_{\odot}$ (see Figure 6 of \cite{m97} 
and \cite{ww95}), while the lower mass progenitors create relatively more
of the heavier $\alpha$- and Fe-group elements.  A possible explanation
for the decline of the light elements without a similar decline in the
heavier elements is to decrease the relative fraction of very massive stars  
in the chemical enrichment history of the $\alpha$-poor stars.
The connection of IMF slope to location in the Galactic potential 
(assuming that the stars with higher \vrf{} formed at greater distances 
from the Galactic center) is unclear.

Another possibility is incomplete mixing.  Using a single Salpeter IMF, 
\cite{arg00} ran simulations of the chemical enrichment history of the 
early Galaxy and included provisions for incomplete mixing.  They found
that incomplete mixing can explain the presence of star-to-star
variations for the light element ratios for the moderately metal-poor stars,
but if the mixing in the halo was less complete at high \vrf, one would
expect to see high-velocity stars with relatively high values of [Na/Fe], 
[Mg/Fe], etc. as well.

        Alternatively, the lower [X/Fe] ratios for some elements may be due to
extra Fe-rich material being added to the gas that formed the stars
that are now at high \vrf{}.  Type Ia supernovae could be the source
of the Fe-group enrichment.  There are, however, a few observational 
details that do not agree with this theory:  

First, other element ratios like [Ca/Fe] and [Ti/Fe] do not show the same 
level of decrease with \vrf{}.  Nucleosynthesis models of Type Ia SN do 
suggest these events create a reasonable amount of Ca and Ti (\cite{n97}), 
but it is unclear whether or not these kind of events can keep 
the mean [Ti/Fe] value approximately constant.  Second, these same 
models predict an overproduction of Ni with respect to Fe by Type Ia SN.  
If anything, the mean [Ni/Fe] ratio in the high velocity stars is lower 
than for the rest of the halo (see Figure 8).

Some of these issues could possibly be resolved by careful analysis of
additional elements.  Oxygen, for example, should be formed mainly by
the very massive Type II supernovae.  That means the [O/Fe] ratio should
track [Na/Fe] and [Mg/Fe].  By a similar argument, [S/Fe] should follow
[Si/Fe].  There may also be clues in the other Fe-group and heavier elements
on the relative contributions of Type Ia and II supernovae.  We are in the
process of analyzing many of these elements in the F00 spectra.

        Another natural next step is to observe more high velocity stars, such 
as those listed in \cite{c88} or elsewhere in the literature.  The
11-star survey of S99 found seven Na-poor stars using kinematic selection
criteria.  These stars are rare, and therefore faint, require time on 8- to 
10-m class telescopes to build up a sizable sample.  However, it is also 
important to continue observing lower velocity stars in order to search for 
Na-poor stars in their midst.  It is paramount to resolve the potential NLTE 
effects in the abundance analysis through either consistent sample selection 
or, preferably, accurate stellar atmosphere modelling.  Finally, it cannot be 
stressed enough that any such new survey needs to follow some sort of 
self-consistent analysis procedure for all the stars, even if it is not the 
same one followed in F00.   Any procedure should also include the analysis of 
a wide sample of stars for comparisons, as for this type of work accurate 
relative abundances are more important than the absolute overall scaling.

\section{Summary of Results}

From the analysis of the self-consistent abundance analysis of F00,
combined with Hipparcos-based kinematics, we have found:

1.  The general trends of the element-to-iron ratios with respect to [Fe/H] are
similar to those seen by previous studies.  

2.  For intermediate-metallicity ($-2 < [Fe/H] < -1$) stars from F00, the 
[X/Fe] ratios for Na, Mg, Al, Si, Ca and Ni, as well as possibly Ti and Y show
a decrease in the value of [X/Fe] with increasing \vrf{}.  

3.  For the metal-poor stars the values of [X/Fe] for the above elements
are correlated.  The correlation between [$\alpha$/Fe] and [Na/Fe] is 
show a higher significance than between any of the individual [X/Fe]
ratios and [Na/Fe]. 
    
4.  When compared to the field halo sample, the dSph giant sample of
\cite{s01} shows a different pattern of abundance ratios than the
field halo sample, including the high \vrf{} stars.  It is unlikely
that the $\alpha$-poor stars in the solar neighborhood originated
the dSph galaxies similar to those studied by \cite{s01}.

\acknowledgments

        The papers of this series make up the PhD thesis of JPF at the 
University of California at Santa Cruz.  JPF wishes
to thank the members of his committee (R. Kraft, R. Peterson, M. Bolte and
P. Gahathakurta) for their efforts and advice.  JPF also wishes to thank
B. Hanson for his assistance with the Hipparcos catalog and the codes for
calculating UVW components for the sample stars, D. Lin for the codes used in 
calculating the orbits of the sample stars and 
J. Johnson and V. Weafer for reviewing drafts of this paper.
Special thanks should got to the anonymous referee for their useful comments.
This research was supported by NSF contract AST 96-18351 to R. P. Kraft
and by the National Research Council of Canada.

%%%%%%%%%%%%%%%%%%%%%%%%%%%%%%
%biblio   references         %
%%%%%%%%%%%%%%%%%%%%%%%%%%%%%%

%%%%%%%%%%%%%%%%%

%%%%%%%%%%%%%%%%%%%%%%%%%
% Figure Captions       %
%%%%%%%%%%%%%%%%%%%%%%%%%

\newpage

\figcaption{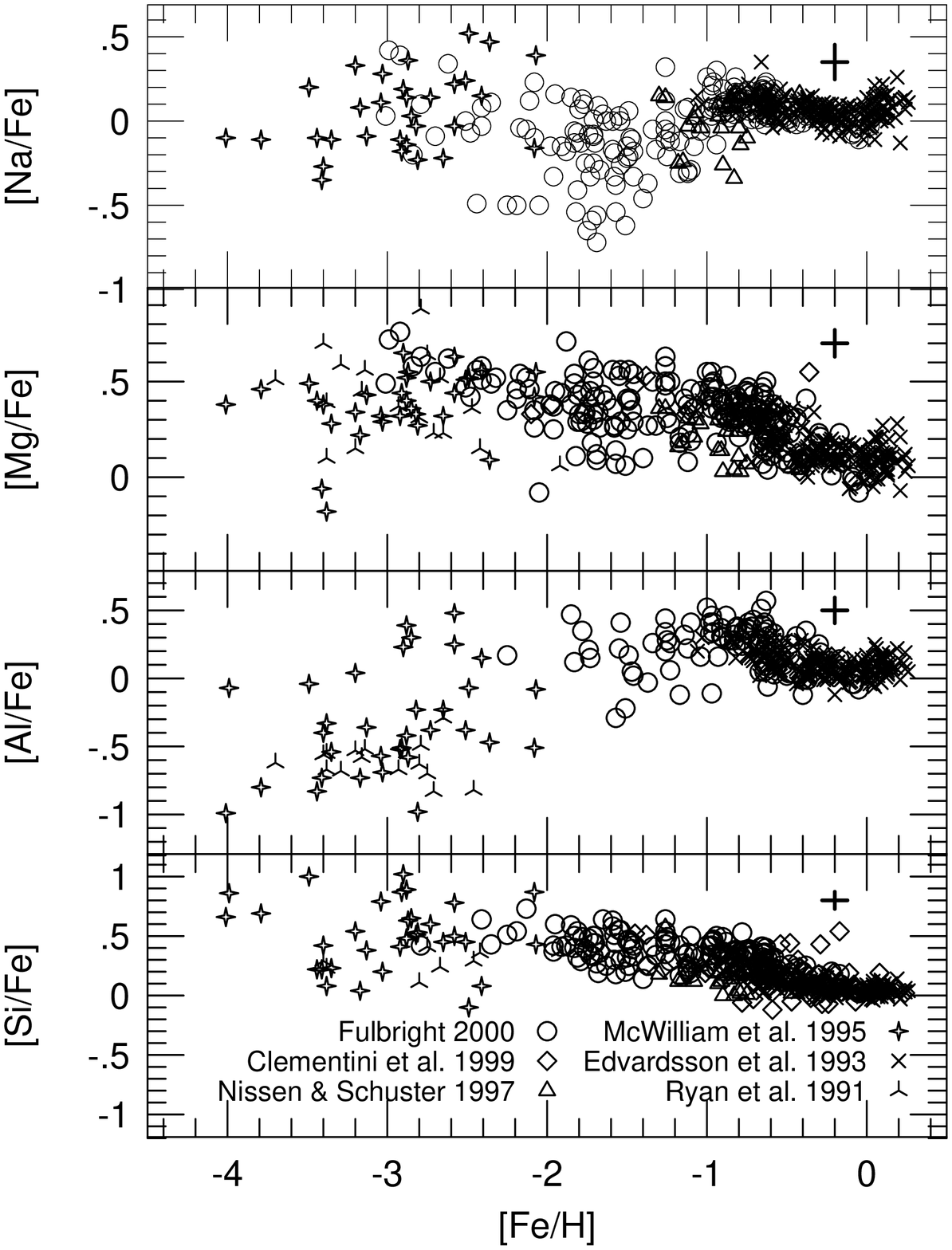}{The distribution of [X/Fe] vs. [Fe/H] for 
Na, Mg, Al and Si for
the F00 and other recent abundance surveys.  The estimated measurement
errors for the F00 data are shown in the upper right.  The F00 star with
low [X/Fe] at [Fe/H] $= -2.05$ for several panels in Figures 1--3 is
the star BD $+80 245$ (= HIP 40068), first studied by \cite{c97}. \label{f1}}

\figcaption{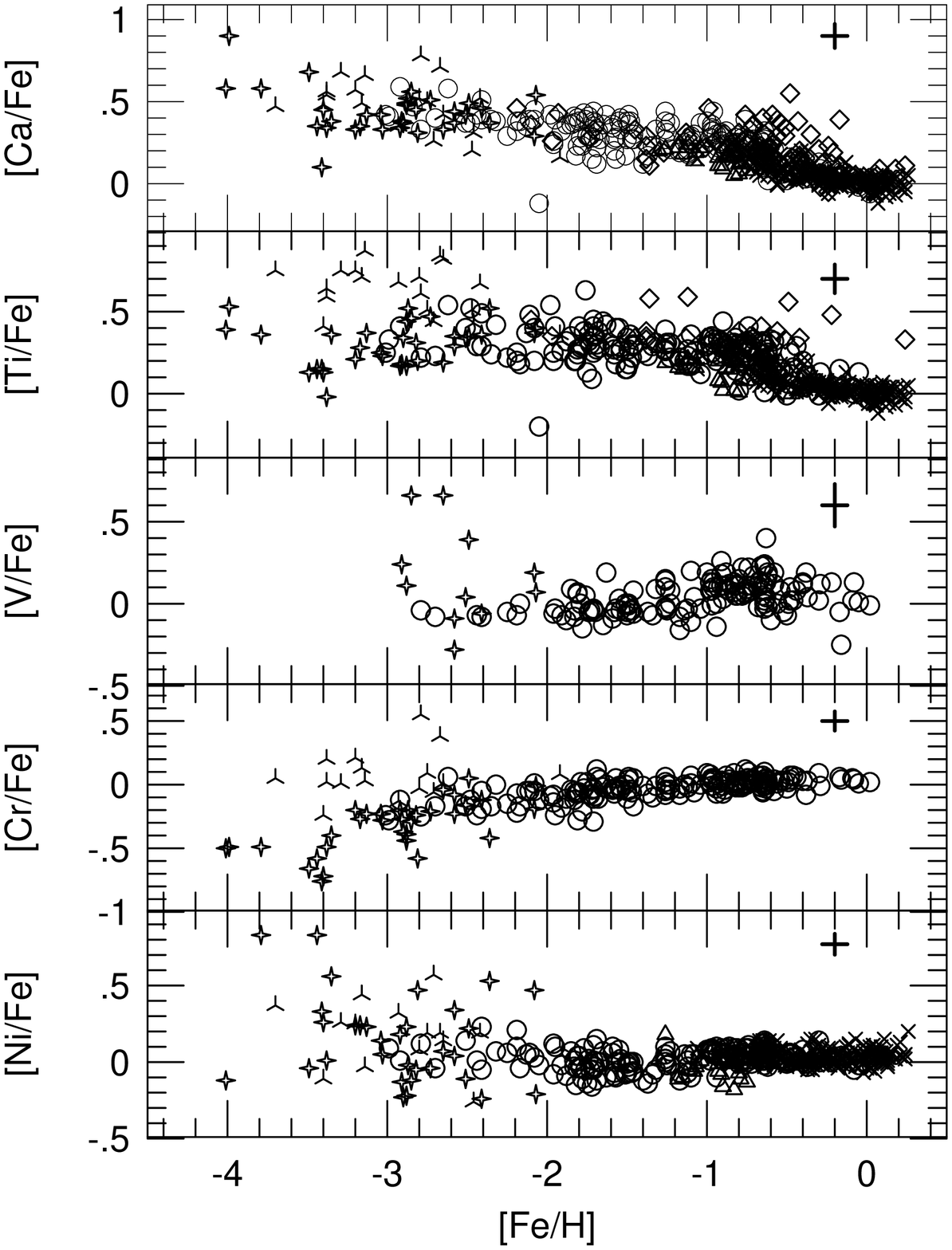}{Same as Figure 1, but for Ca, Ti, V, Cr and 
Ni. \label{f2}}

\figcaption{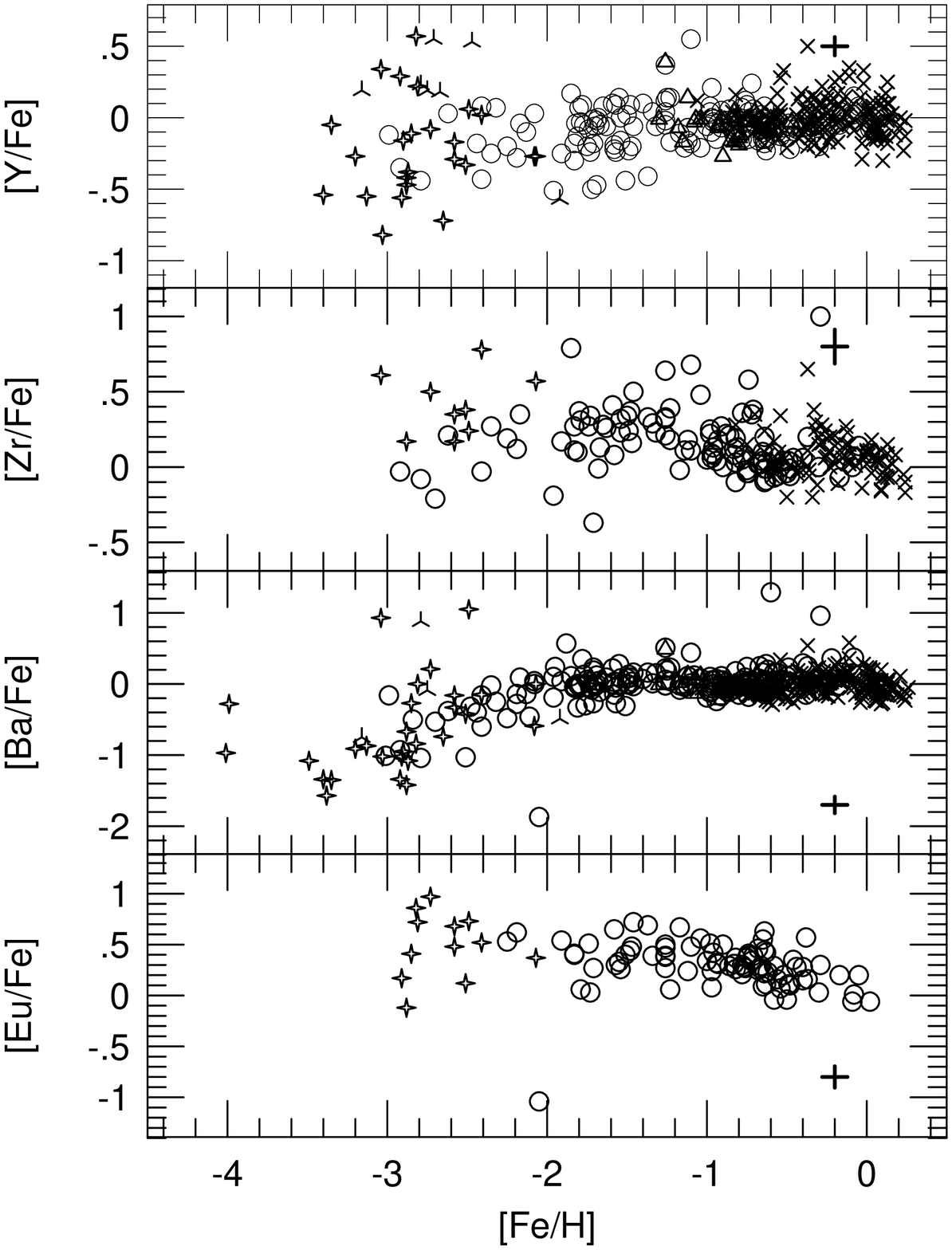}{Same as Figure 1, but for Y, Zr, Ba and Eu. 
As mentioned in F00, the [Eu/Fe] measurement for BD $+80 245$ is an upper 
limit. \label{f3}}

\figcaption{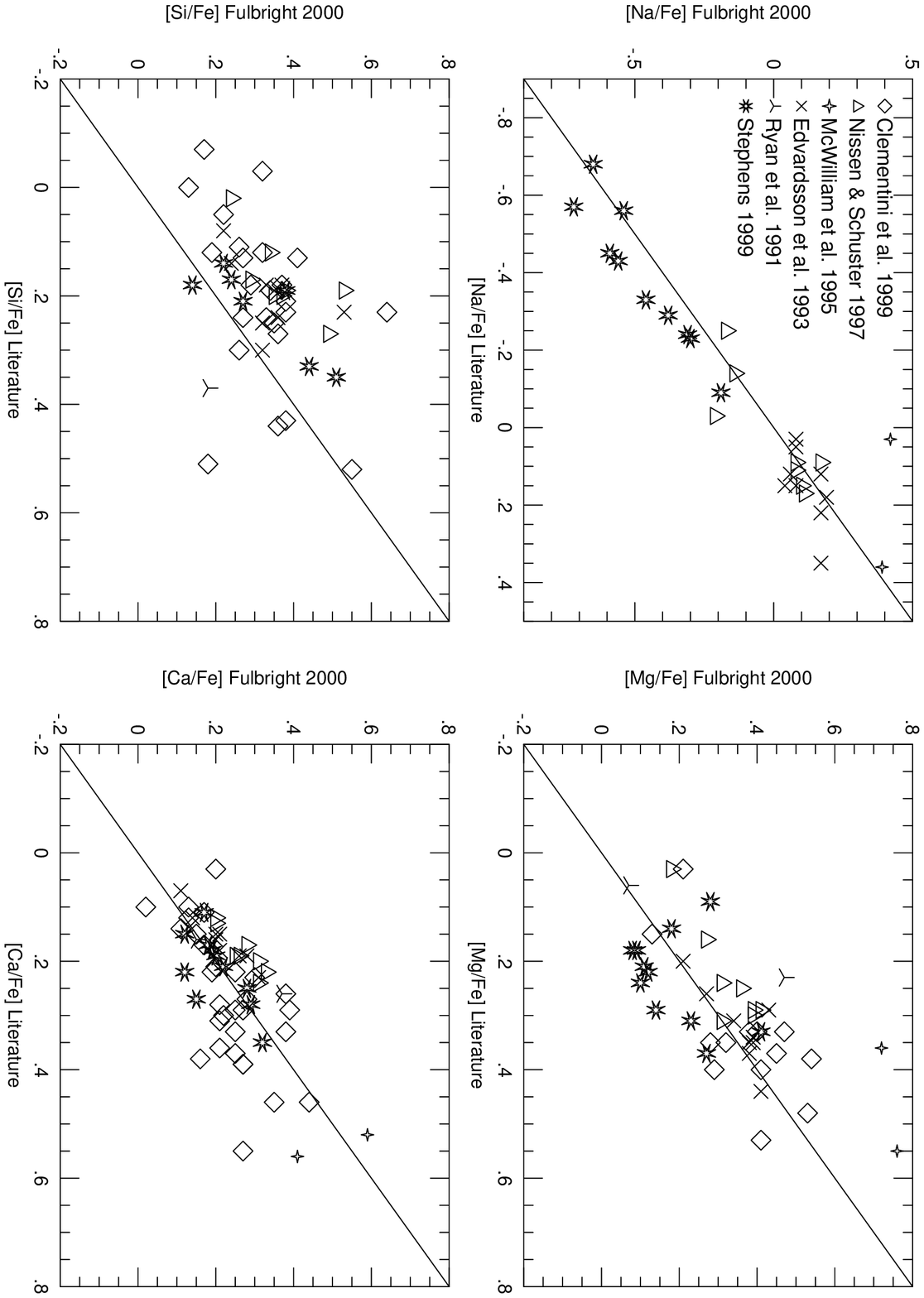}{Comparison of the measured abundance ratios 
for stars in common
between F00 and the other abundance surveys included in Figures 1--3.  As
discussed in the text, the [Si/Fe] ratio shows a $\sim 0.1$ dex offset
between the F00 and other surveys. \label{f4}}

\figcaption{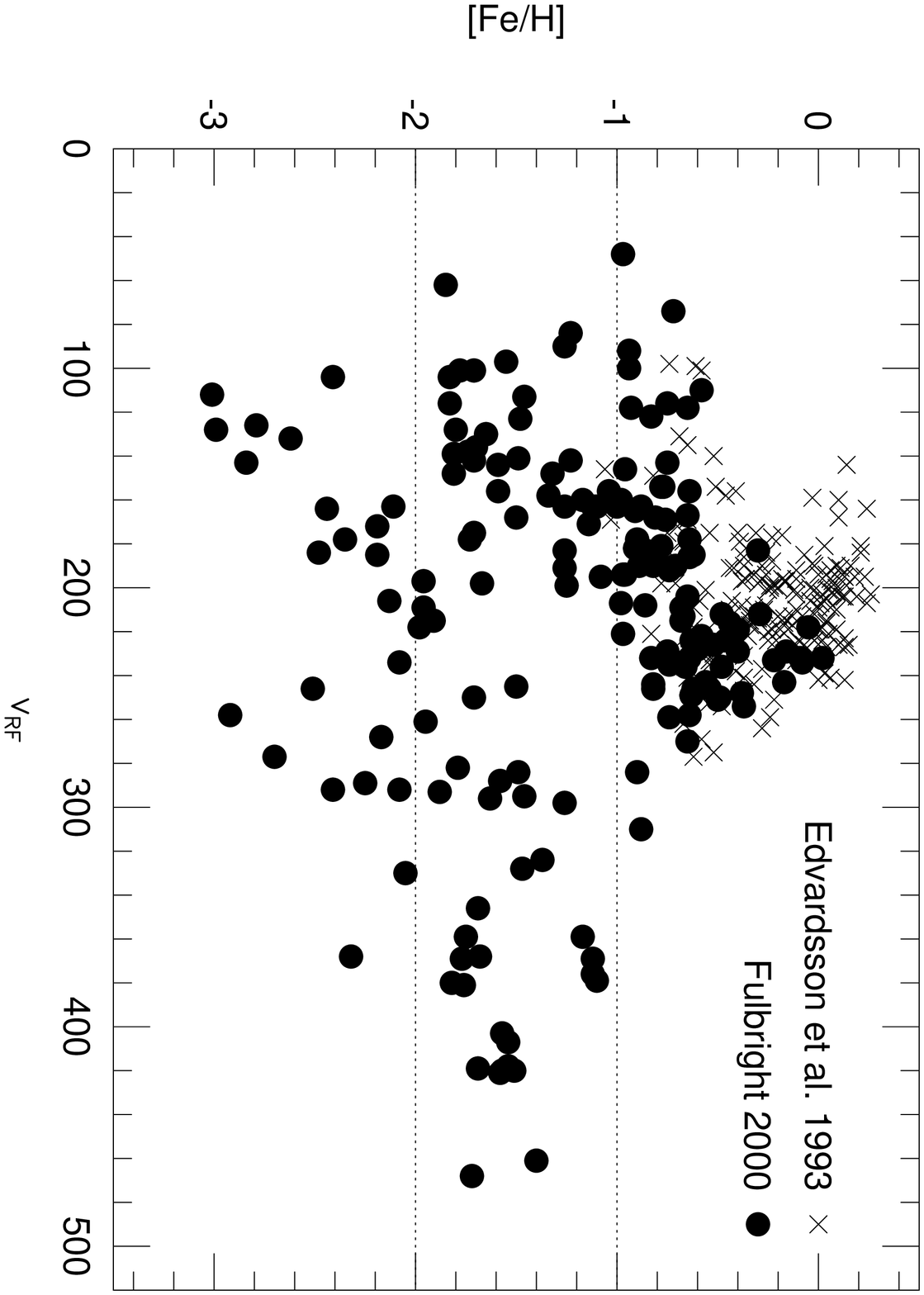}{The distribution of [Fe/H] vs. \vrf{} for the 
F00 data.  There is
a lack of very metal-poor ([Fe/H] $< -2$) high velocity stars, although such
stars are known to exist.  For the metal-rich ([Fe/H] $> -1$) stars, the F00
and \cite{edv93} disk stars cover the same region, suggesting that the more
metal-rich F00 stars are members of the disk population.  The dotted lines 
denote the intermediate metallicity ``selected sample.'' \label{f5}}

\figcaption{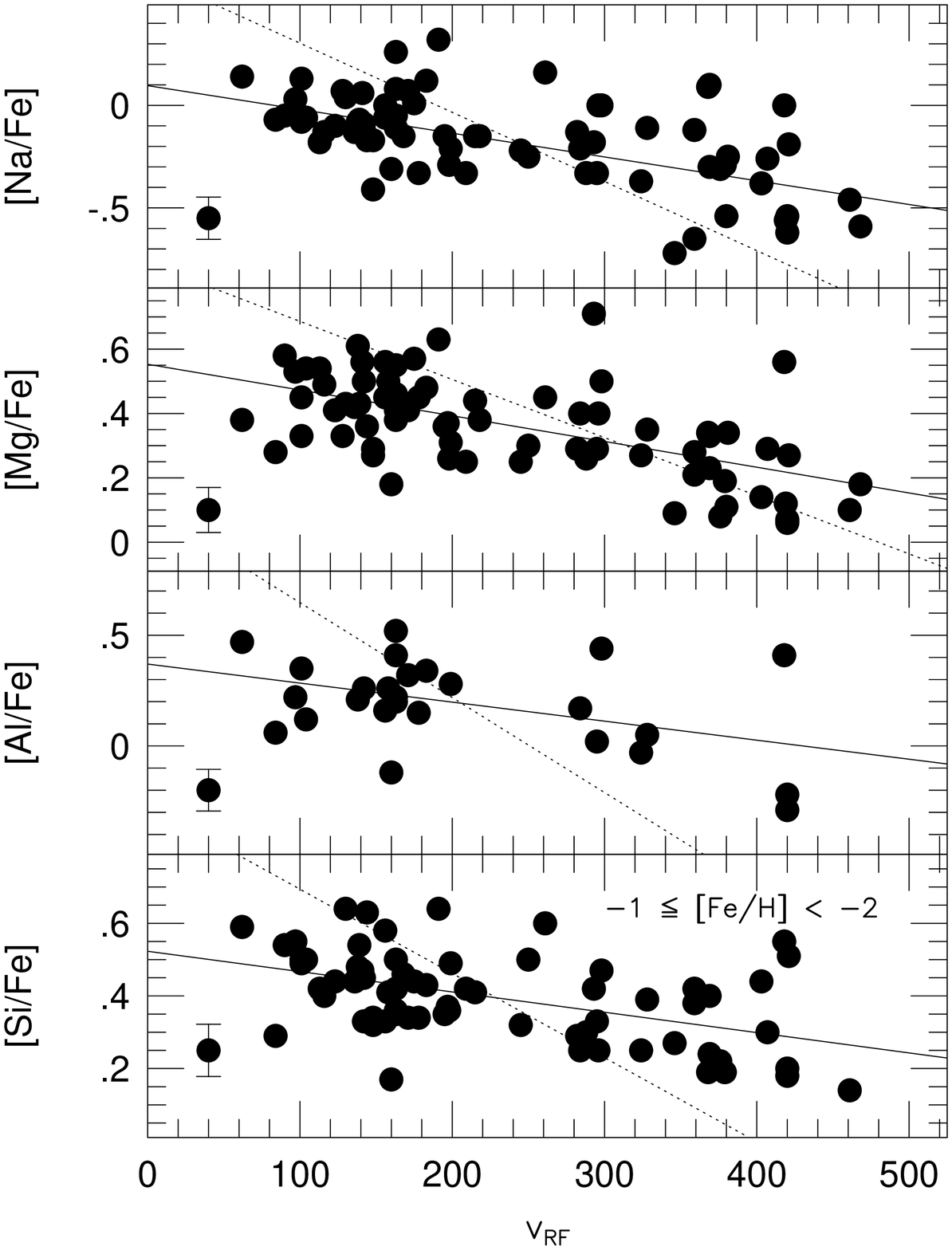}{Plots of the distribution of [X/Fe] vs. \vrf{} 
for Na, Mg, Al 
and Si for the selected sample.  All four of these distributions show 
statistically significant correlations with respect to \vrf{}.  The solid
line is the least-squares fit to the data, while the dotted line is the
reverse least-squares fit.  The estimated measurement uncertainty for the
abundance ratios is shown in the lower left. \label{f6}}

\figcaption{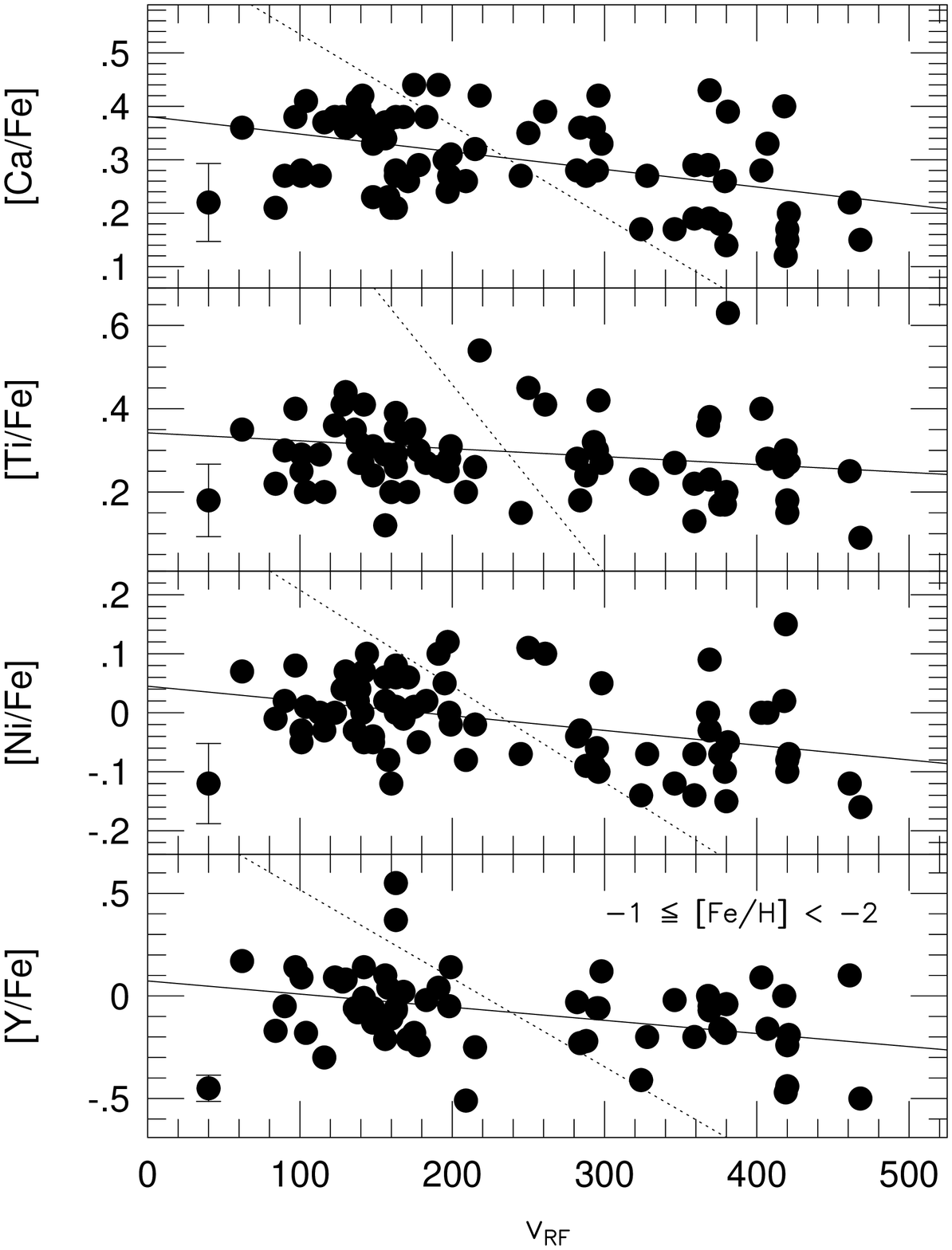}{Same as Figure 6, but for Ca, Ti, Ni and Y.  
The correlations 
for Ti and Y are not as significant as the other six elements.  The other
elements studied in F00 (V, Cr, Zr, Ba and Eu) do not show any statistically
significant correlations with \vrf{}. \label{f7}}

\figcaption{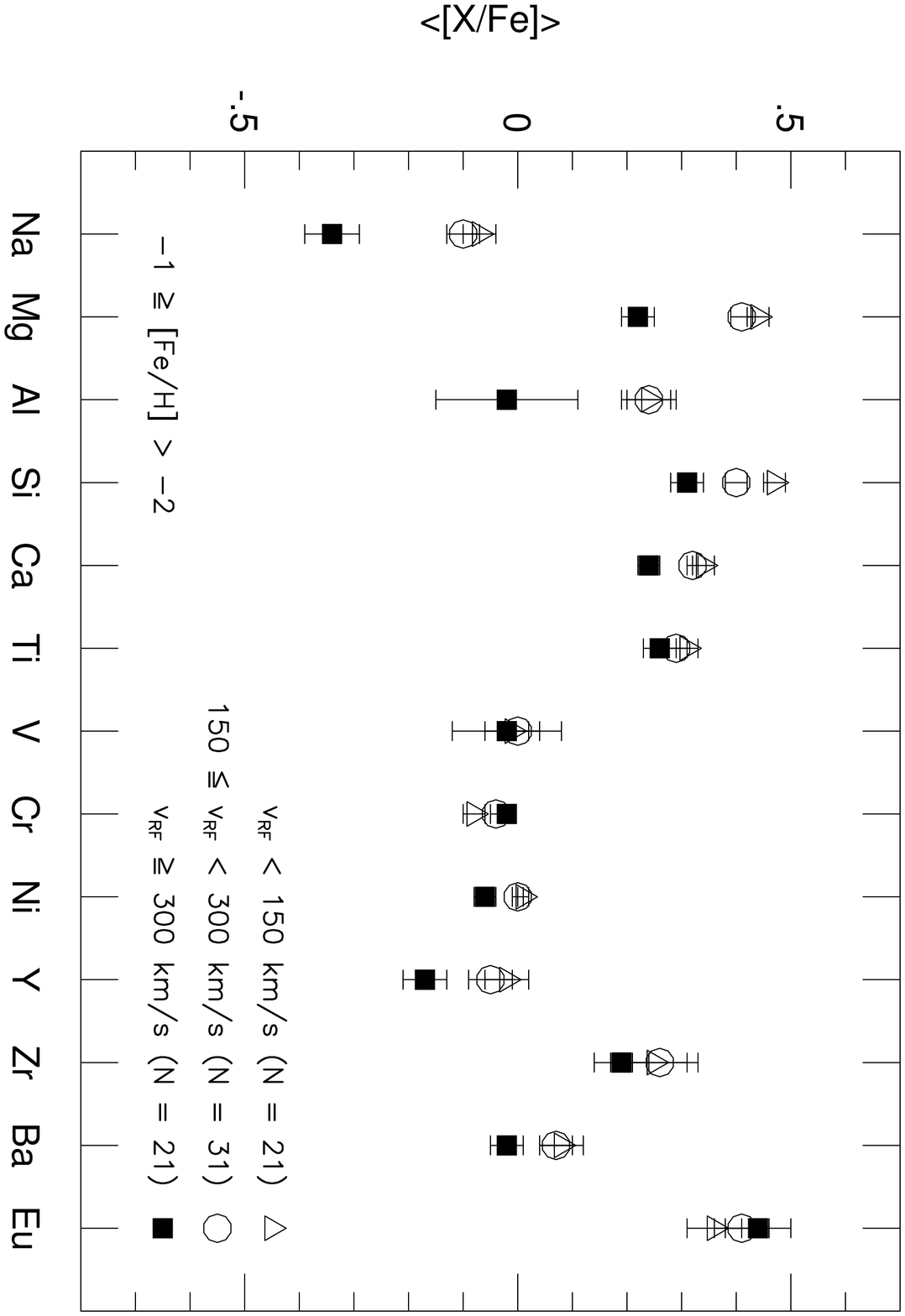}{Figure 8.  Plots of the mean value of [X/Fe] for the selected sample broken
into three \vrf{} groups.  The highest velocity group has significantly lower
mean values for several elements.  Note that the number of measurements for 
any given ratio may be smaller than the number of stars in that group.  This
is especially a problem for Al, which helps explain the large sdom error
bars for that ratio.   \label{f8}}

\figcaption{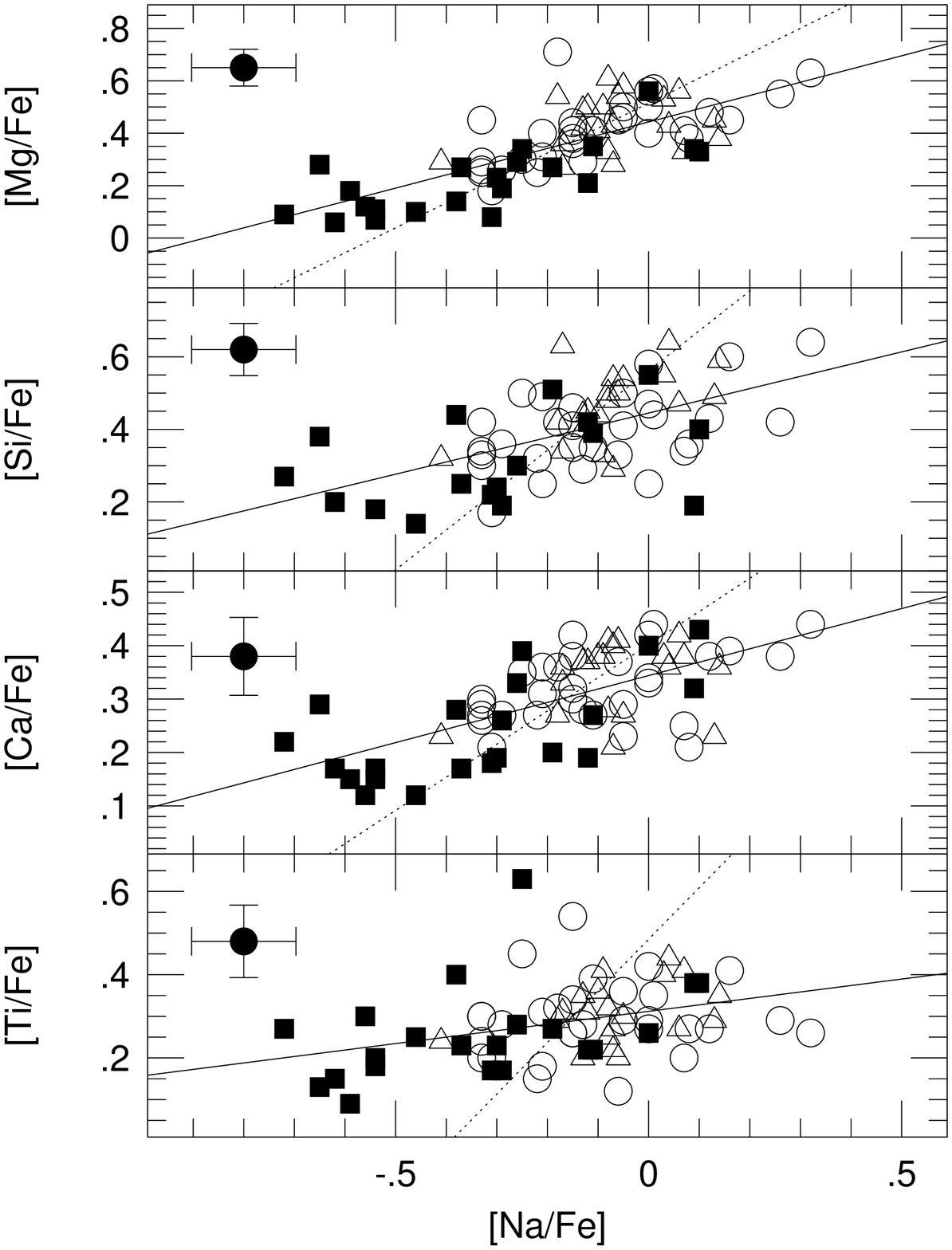}{Plots of the correlation between the [X/Fe] 
ratios for the four $\alpha$-elements measured in F00 and [Na/Fe].  The
symbols are the same as in Figure 8, and the error bars denote the
estimated measurement uncertainty for the ratios.  As in Figures 6 and 7,
the solid line marks the least-squares fit to the data, while the dotted line
is the reverse fit.  The displacement of the high velocity stars to lower
[X/Fe] values is clear in these plots. \label{f9}}

\figcaption{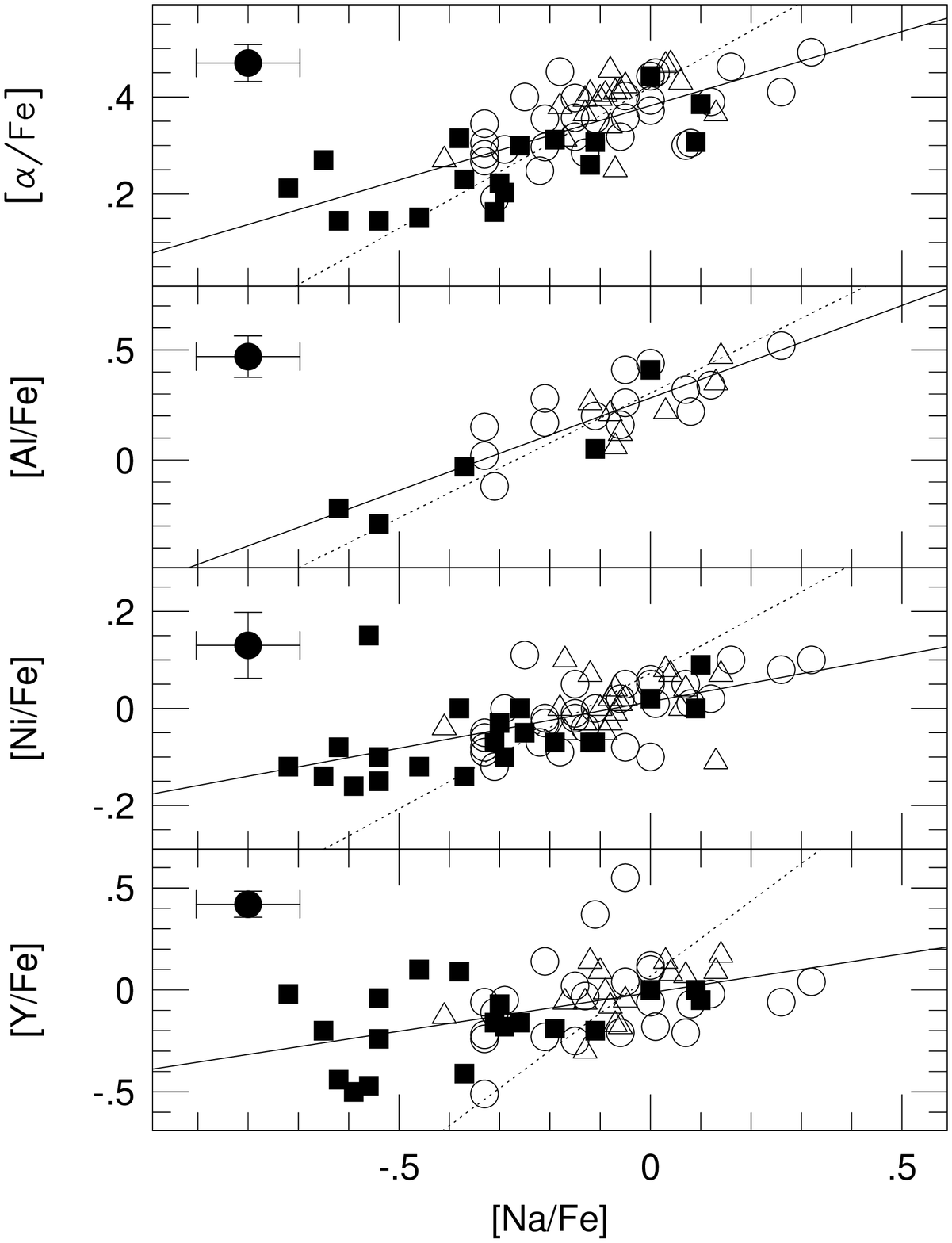}{Same as Figure 9, but for the mean 
$\alpha$-element abundance
(for only the 65 stars with measurements of all four elements), Al, Ni,
and Y.  The statistical significance of the [$\alpha$/Fe] vs. [Na/Fe] 
correlation is greater than any of the four individual $\alpha$-elements
with [Na/Fe].   \label{f10}}

\figcaption{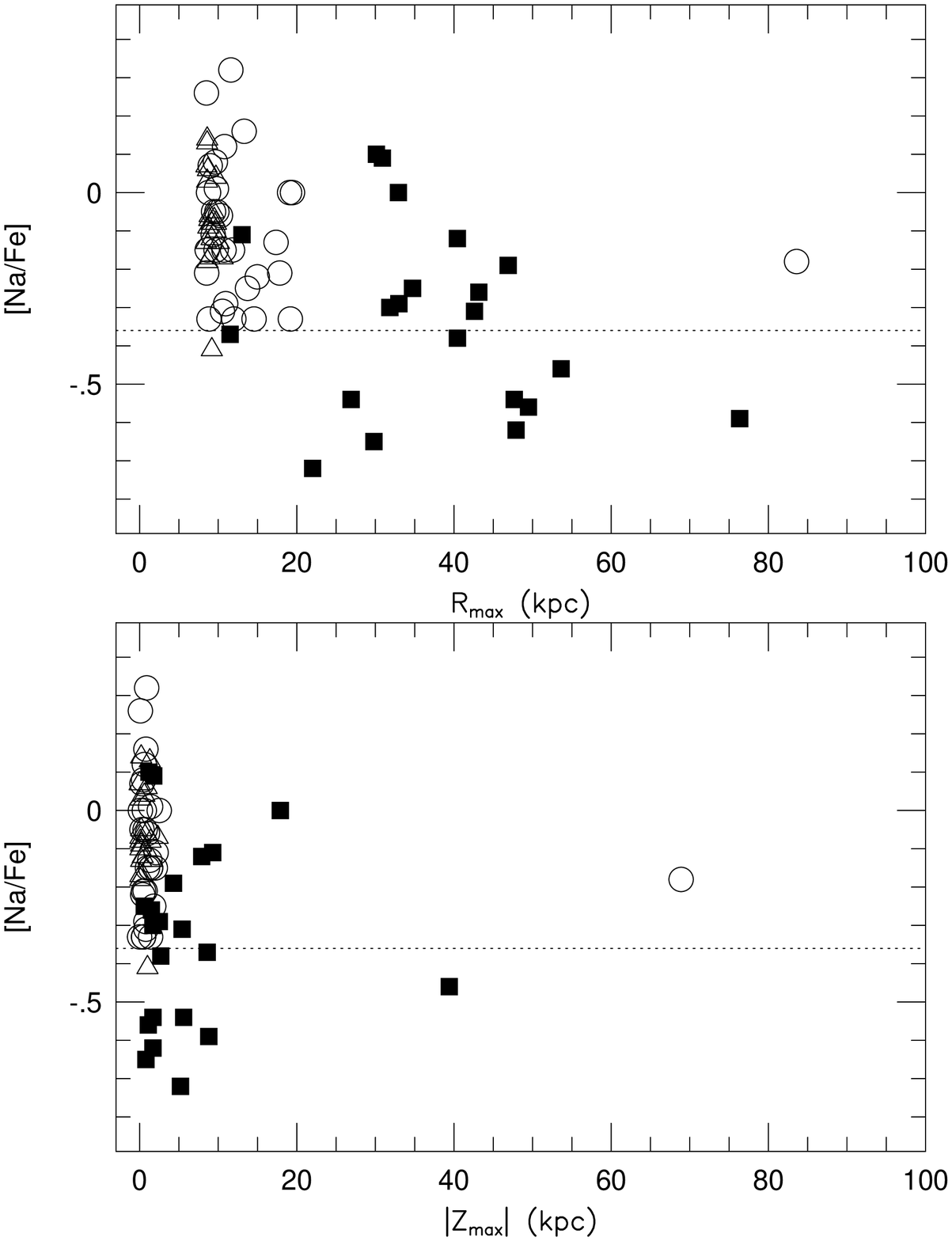}{The value of [Na/Fe] as a function of 
$R_{max}$ and $|Z_{max}|$
for the selected sample.  The symbols are the same as in Figure 8, and the 
dotted line denotes the mean [Na/Fe] ratio of the highest \vrf{} group 
([Na/Fe] $= -0.36$).  Approximately half of the stars with $R_{max} > 20$ kpc 
or $|Z_{max}| > 5$ kpc show [Na/Fe] ratios lower than nearly all of the lower 
\vrf{} halo stars. \label{f11}}

\figcaption{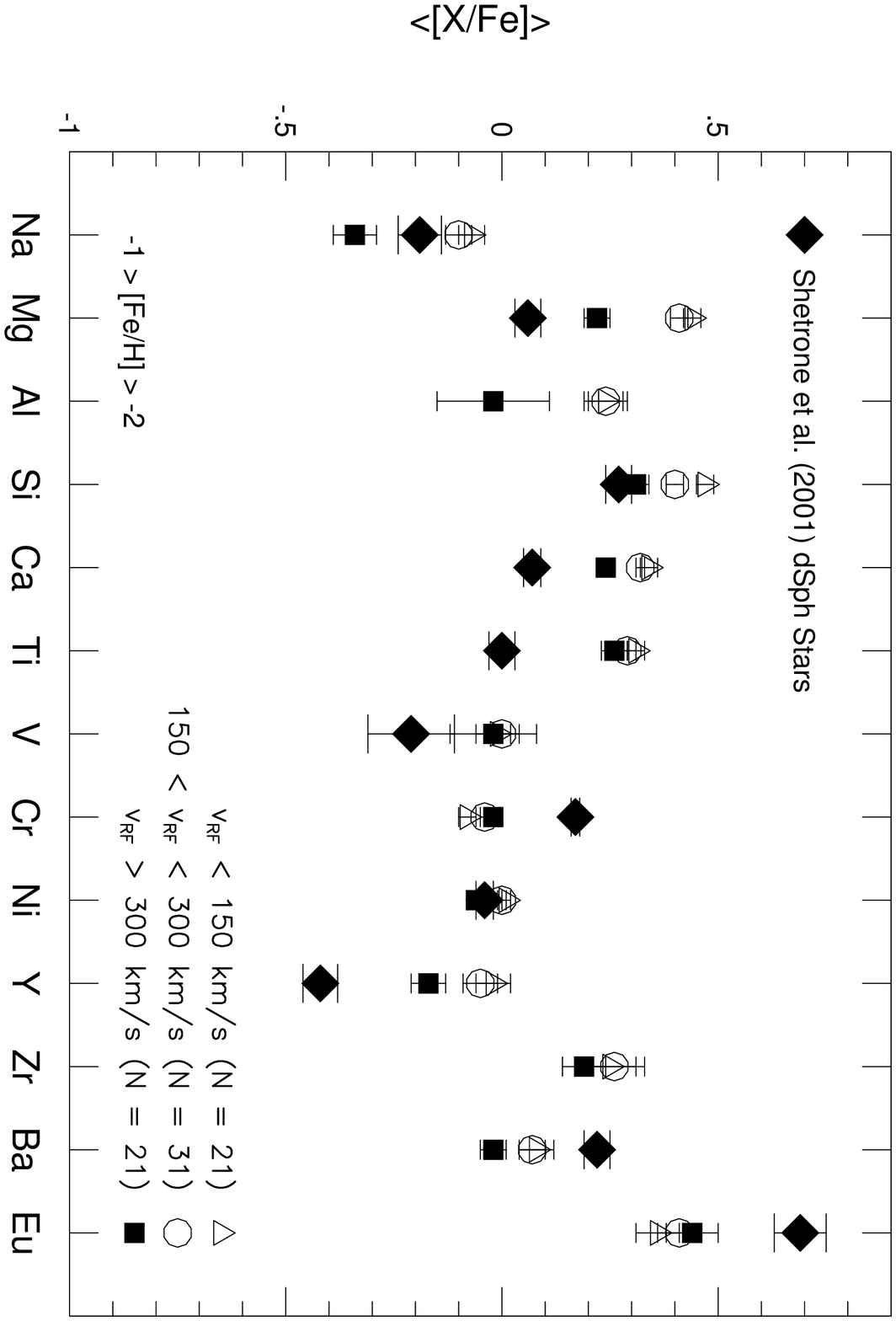}{Same as Figure 8, but with the mean [X/Fe] 
ratios for the 
intermediate metallicity dSph stars from \cite{s01} plotted as well.
While the dSph stars show similar mean values to the high velocity
stars for some elements, the large differences seen in the mean ratios
for Ca, Ti, Cr, Y, Ba and Eu strongly suggest that the high velocity
stars were not accreted from disrupted dSph galaxies like the ones
observed by \cite{s01}. \label{f12}}

%%%%%%%%%%%%%%%%%%%%%%%%%
% Figures               %
%%%%%%%%%%%%%%%%%%%%%%%%%

\plotone{Fulbright.fig1.ps}

\plotone{Fulbright.fig2.ps}

\plotone{Fulbright.fig3.ps}

\plotone{Fulbright.fig4.ps}

\plotone{Fulbright.fig5.ps}

\plotone{Fulbright.fig6.ps}

\plotone{Fulbright.fig7.ps}

\plotone{Fulbright.fig8.ps}

\plotone{Fulbright.fig9.ps}

\plotone{Fulbright.fig10.ps}

\plotone{Fulbright.fig11.ps}

\plotone{Fulbright.fig12.ps}

%%%%%%%%%%%%%%%%%%%%%%%%%
% Tables                %
%%%%%%%%%%%%%%%%%%%%%%%%%

%\renewcommand{\arraystretch}{.6}

%%%%%%%%%%%%
% Table 1  %
%%%%%%%%%%%%
\clearpage
\begin{deluxetable}{rrrrrrrrrrrrr}
\tablenum{1}
\tiny 
\tablewidth{0pt}
\tablecaption{Kinematic and Orbital Data}
\tablehead{
\colhead{HIP\tablenotemark{a}}&\colhead{RV}&\colhead{Ref\tablenotemark{b}}&\colhead{$U_{LSR}$}&\colhead{$V_{LSR}$}&\colhead{$W_{LSR}$}&\colhead{$v_{RF}$}&\colhead{$v_{ROT}$}&\colhead{h} & \colhead {$R_{min}$} & \colhead{$R_{max}$} & \colhead{$e$} & \colhead{$|Z_{max}|$}\\
 &\colhead{km/s}& & \colhead{km/s} &\colhead{km/s} &\colhead{km/s} &\colhead{km/s} & \colhead{km/s} & \colhead{km/s} &\colhead{kpc} &\colhead{kpc} & &\colhead{kpc} }
\startdata
   SUN & \nodata & \nodata & $  -9$	& $  12$ & $   7$ & $232$ & $ 232$ &$ 232$& 8.4 & 8.9 & 0.03 & 0.1  \nl
   171 &$ -36$&1 &$  -1$ & $ -59$ & $ -23$ & $163$ & $ 161$ &$ 163$& 4.7 & 8.5 & 0.29 & 0.2  \nl
  2413 &$-378$&4 &$-161$ & $-345$ & $ -46$ & $209$ & $-125$ &$-133$ & 2.7 & 12.0 & 0.64 & 0.5  \nl
  3026 &$ -49$&2 &$-144$ & $-223$ & $ -34$ & $148$ & $  -3$ &$  34$ & 0.0 & 10.4 & 0.99 & 0.4  \nl
  3086 &$ -28$&2 &$ 156$ & $ -42$ & $  55$ & $243$ & $ 178$ &$ 186$ & 4.3 & 12.8 & 0.49 & 0.7  \nl
  3554 &$ -67$&1 &$ 177$ & $ -55$ & $  86$ & $258$ & $ 165$ &$ 187$ & 3.9 & 14.1 & 0.57 & 2.2  \nl
  5336 &$ -24$&2 &$  96$ & $ -41$ & $ -41$ & $207$ & $ 179$ &$ 183$ & 4.3 & 10.2 & 0.35 & 0.4  \nl
  5445 &$-321$&1 &$ 118$ & $-468$ & $  89$ & $288$ & $-248$ &$-263$ & 6.9 & 14.6 & 0.36 & 1.4  \nl
  5458 &$-245$&1 &$-131$ & $-192$ & $  81$ & $156$ & $  28$ &$  85$ & 0.5 & 10.3 & 0.91 & 1.0  \nl
  6710 &$-110$&1 &$ -14$ & $-300$ & $ -64$ & $104$ & $ -80$ &$-103$ & 1.8 & 8.9 & 0.66 & 1.0  \nl
  7217 &$ -53$&2 &$-102$ & $  -8$ & $  -9$ & $236$ & $ 212$ &$ 212$ & 6.0 & 11.3 & 0.30 & 0.1  \nl
 10140 &$  27$&2 &$  52$ & $ -63$ & $ -41$ & $171$ & $ 157$ &$ 162$ & 4.4 & 9.0 & 0.34 & 0.4  \nl
 10449 &$  28$&2 &$ 194$ & $-186$ & $  64$ & $207$ & $  34$ &$  72$ & 0.5 & 12.4 & 0.92 & 0.8    \nl
 11349 &$ -11$&2 &$  19$ & $ -12$ & $  37$ & $212$ & $ 208$ &$ 211$ & 7.1 & 8.7  & 0.10 & 0.3  \nl
 11952 &$  24$&2 &$ -38$ & $ -86$ & $ -30$ & $142$ & $ 134$ &$ 137$ & 3.5 & 8.7 & 0.42 & 0.3  \nl
 12306 &$-105$&5 &$-146$ & $ -21$ & $ -30$ & $249$ & $ 199$ &$ 202$ & 5.1 & 12.9 & 0.44 & 0.3  \nl
 13366 &$   6$&2 &$ -46$ & $ -93$ & $ -74$ & $154$ & $ 127$ &$ 147$ & 3.3 & 8.8 & 0.45 & 0.8  \nl
 14086 &$  42$&3 &$ -17$ & $ -71$ & $ -14$ & $151$ & $ 149$ &$ 159$ & 4.2 & 8.6 & 0.34 & 0.1  \nl
 14594 &$-140$&2 &$-165$ & $-110$ & $ -58$ & $206$ & $ 110$ &$ 124$ & 2.4 & 11.7 & 0.67 & 0.7  \nl
 15394 &$   8$&2 &$ -26$ & $ -45$ & $ -49$ & $183$ & $ 175$ &$ 181$ & 5.4 & 8.7 & 0.24 & 0.4  \nl
 16214 &$ 151$&4 &$  84$ & $-281$ & $ -92$ & $138$ & $ -61$ &$-110$ & 1.2 & 9.7 & 0.77 & 1.5  \nl
 17085 &$  -5$&3 &$ -13$ & $  13$ & $  10$ & $233$ & $ 233$ &$ 233$ & 8.4 & 9.1 & 0.04 & 0.1  \nl
 17147 &$ 120$&2 &$ 100$ & $ -73$ & $ -38$ & $182$ & $ 147$ &$ 152$ & 3.8 & 9.9 & 0.45 & 0.3  \nl
 17666 &$  51$&2 &$  86$ & $-100$ & $ -68$ & $163$ & $ 120$ &$ 138$ & 2.9 & 9.4 & 0.52 & 0.7  \nl
 18235 &$  98$&1 &$  17$ & $-150$ & $ -18$ & $ 74$ & $  70$ &$  72$ & 1.4 & 8.6 & 0.72 & 0.1  \nl
 18915 &$ -26$&2 &$  30$ & $-173$ & $  27$ & $ 62$ & $  47$ &$  54$ & 0.8 & 8.6 & 0.82 & 0.2  \nl
 18995 &$-112$&1 &$ -80$ & $-257$ & $  19$ & $ 90$ & $ -37$ &$ -42$ & 0.6 & 9.4 & 0.88 & 0.2  \nl
 19007 &$  37$&3 &$  43$ & $  23$ & $   7$ & $246$ & $ 243$ &$ 243$ & 7.9 & 10.5 & 0.14 & 0.1  \nl
 19378 &$  14$&1 &$  22$ & $ -51$ & $  49$ & $178$ & $ 169$ &$ 176$ & 5.3 & 8.8 & 0.25 & 0.6  \nl
 19797 &$ 345$&5 &$ 349$ & $-130$ & $ -76$ & $368$ & $  90$ &$ 117$ & 1.4 & 30.9 & 0.91 & 1.8  \nl
 21000 &$  45$&3 &$  34$ & $   6$ & $  -7$ & $229$ & $ 226$ &$ 227$ & 7.6 & 9.4 & 0.10 & 0.1  \nl
 21586 &$  41$&3 &$  67$ & $ -51$ & $  -3$ & $182$ & $ 169$ &$ 169$ & 4.8 & 9.3 & 0.32 & 0.0  \nl
 21609 &$  56$&1 &$ 368$ & $-126$ & $  35$ & $381$ & $  94$ &$ 100$ & 1.5 & 34.7 & 0.92 & 0.7  \nl
 21648 &$  24$&1 &$-228$ & $-164$ & $-176$ & $293$ & $  56$ &$ 184$ & 0.9 & 83.6 & 0.98 & 68.9  \nl
 21767 &$  -3$&3 &$  -7$ & $  -4$ & $  16$ & $216$ & $ 216$ &$ 216$ & 7.7 & 8.6 & 0.06 & 0.1  \nl
 22246 &$  96$&2 &$  95$ & $   7$ & $  34$ & $248$ & $ 227$ &$ 229$ & 6.5 & 11.7 & 0.28 & 0.3  \nl
 22632 &$ 111$&6 &$  54$ & $ -88$ & $ -23$ & $144$ & $ 132$ &$ 134$ & 3.4 & 8.9 & 0.44 & 0.2  \nl
 23344 &$ 174$&2 &$ 124$ & $-155$ & $ -28$ & $143$ & $  65$ &$  71$ & 1.2 & 10.1 & 0.78 & 0.2  \nl
 24316 &$ 232$&1 &$ 198$ & $-335$ & $ 102$ & $250$ & $-115$ &$-154$ & 2.4 & 13.7 & 0.70 & 1.8  \nl
 26688 &$   2$&3 &$ -42$ & $   1$ & $ -35$ & $228$ & $ 221$ &$ 224$ & 7.3 & 9.5 & 0.13 & 0.3  \nl
 27654 &$  99$&1 &$ -26$ & $-130$ & $ -35$ & $100$ & $  90$ &$  97$ & 2.0 & 8.6 & 0.62 & 0.3  \nl
 28188 &$  39$&3 &$ 160$ & $ -32$ & $  -5$ & $247$ & $ 188$ &$ 188$ & 4.6 & 13.3 & 0.49 & 0.0  \nl
 29759 &$ 242$&2 &$ 264$ & $-251$ & $ -28$ & $268$ & $ -31$ &$ -41$ & 0.4 & 17.0 & 0.95 & 0.3  \nl
 29992 &$ 167$&1 &$  75$ & $-154$ & $  18$ & $101$ & $  66$ &$  68$ & 1.3 & 9.1 & 0.75 & 0.1  \nl
 30668 &$ 307$&6 &$ 238$ & $-175$ & $ -36$ & $245$ & $  45$ &$  58$ & 0.7 & 15.0 & 0.91 & 0.4  \nl
 30990 &$  60$&2 &$  77$ & $ -52$ & $  45$ & $190$ & $ 168$ &$ 174$ & 4.7 & 9.6 & 0.34 & 0.4  \nl
 31188 &$  47$&7 &$  17$ & $ -45$ & $  66$ & $188$ & $ 175$ &$ 187$ & 5.5 & 8.6 & 0.22 & 0.7  \nl
 31639 &$  53$&3 &$  -6$ & $ -39$ & $ -38$ & $185$ & $ 181$ &$ 185$ & 5.7 & 8.5 & 0.20 & 0.3  \nl
 32308 &$ 105$&3 &$  78$ & $ -61$ & $   0$ & $178$ & $ 159$ &$ 159$ & 4.4 & 9.5 & 0.37 & 0.0  \nl
 33582 &$ -94$&2 &$-212$ & $-120$ & $  12$ & $235$ & $ 100$ &$ 101$ & 1.9 & 13.9 & 0.76 & 0.1  \nl
 34146 &$ -14$&3 &$ -39$ & $   5$ & $   2$ & $229$ & $ 225$ &$ 225$ & 7.5 & 9.6 & 0.12 & 0.0  \nl
 34548 &$ -28$&3 &$ -37$ & $   0$ & $   6$ & $224$ & $ 220$ &$ 220$ & 7.4 & 9.4 & 0.11 & 0.1  \nl
 36491 &$  91$&2 &$  49$ & $-113$ & $   0$ & $118$ & $ 107$ &$ 107$ & 2.5 & 8.8 & 0.55 & 0.0  \nl
 36849 &$ -34$&2 &$ -67$ & $ -78$ & $ -46$ & $163$ & $ 142$ &$ 149$ & 3.8 & 9.2 & 0.42 & 0.4  \nl
 37335 &$  66$&5 &$-170$ & $-268$ & $  71$ & $191$ & $ -48$ &$ -86$ & 0.9 & 11.6 & 0.86 & 0.9  \nl
 38541 &$-235$&2 &$-270$ & $-206$ & $ -82$ & $282$ & $  14$ &$  83$ & 0.2 & 17.4 & 0.98 & 1.4  \nl
 38621 &$-108$&8 &$  20$ & $-159$ & $-124$ & $139$ & $  61$ &$ 138$ & 1.3 & 9.5 & 0.76 & 2.4  \nl
 38625 &$ 102$&2 &$  73$ & $ -53$ & $   8$ & $183$ & $ 167$ &$ 167$ & 4.7 & 9.4 & 0.34 & 0.1  \nl
 40068 &$   4$&2 &$ 178$ & $-352$ & $ 244$ & $330$ & $-132$ &$-278$ & 5.1 & 18.0 & 0.56 & 10.5  \nl
 40778 &$  66$&2 &$ 130$ & $-246$ & $  32$ & $136$ & $ -26$ &$ -41$ & 0.4 & 10.1 & 0.92 & 0.3  \nl
 42592 &$ 206$&2 &$-253$ & $-349$ & $  68$ & $292$ & $-129$ &$-146$ & 2.6 & 18.0 & 0.75 & 1.1  \nl
 44075 &$ 121$&1 &$  40$ & $ -80$ & $  77$ & $165$ & $ 140$ &$ 160$ & 3.8 & 8.8 & 0.39 & 0.8  \nl
 44116 &$  17$&3 &$ -18$ & $  -2$ & $  36$ & $222$ & $ 218$ &$ 221$ & 7.7 & 8.8 & 0.07 & 0.3  \nl
 44124 &$  35$&3 &$-191$ & $-172$ & $  14$ & $197$ & $  48$ &$  50$ & 0.8 & 12.2 & 0.88 & 0.1  \nl
 44716 &$ 170$&2 &$ 119$ & $-355$ & $ -75$ & $195$ & $-135$ &$-154$ & 3.3 & 10.7 & 0.53 & 1.1  \nl
 44919 &$ -36$&9 &$ -60$ & $  42$ & $  30$ & $270$ & $ 262$ &$ 263$ & 7.8 & 12.6 & 0.23 & 0.3  \nl
 47139 &$ 110$&8 &$  18$ & $-108$ & $  -2$ & $113$ & $ 112$ &$ 112$ & 2.8 & 8.7 & 0.51 & 0.2  \nl
 47640 &$ -14$&3 &$ -17$ & $  13$ & $  -5$ & $234$ & $ 233$ &$ 233$ & 8.3 & 9.2 & 0.05 & 0.1  \nl
 48146 &$  -2$&3 &$ -23$ & $  -4$ & $   2$ & $218$ & $ 216$ &$ 216$ & 7.5 & 8.9 & 0.09 & 0.1  \nl
 48152 &$ -15$&3 &$-234$ & $-225$ & $  -0$ & $234$ & $  -5$ &$  -5$ & 0.1 & 14.3 & 0.99 & 0.1  \nl
 49371 &$ -40$&7 &$-134$ & $  -4$ & $  60$ & $261$ & $ 216$ &$ 224$ & 5.8 & 13.3 & 0.39 & 0.9  \nl
 50139 &$ -22$&2 &$ -80$ & $ -21$ & $ -15$ & $215$ & $ 199$ &$ 199$ & 5.8 & 10.1 & 0.27 & 0.1  \nl
 50173 &$ -36$&1 &$ -39$ & $-117$ & $  20$ & $112$ & $ 103$ &$ 105$ & 2.6 & 9.2 & 0.56 & 0.7  \nl
 52771 &$  74$&1 &$-163$ & $-328$ & $  95$ & $218$ & $-108$ &$-144$ & 2.4 & 11.8 & 0.67 & 1.5  \nl
 53070 &$  65$&2 &$  26$ & $-128$ & $  18$ & $ 97$ & $  92$ &$  93$ & 2.1 & 8.6 & 0.61 & 0.1  \nl
 54858 &$ -22$&11&$ 317$ & $-303$ & $-147$ & $359$ & $ -83$ &$-168$ & 1.5 & 40.4 & 0.93 & 8.0  \nl
 55022 &$  62$&2 &$-115$ & $  18$ & $ 102$ & $284$ & $ 238$ &$ 259$ & 6.6 & 13.9 & 0.36 & 1.8  \nl
 57265 &$ 198$&2 &$ 367$ & $-214$ & $  94$ & $379$ & $	6$ &$  95$ & 0.1 & 33.0 & 0.99 & 2.6  \nl
 57850 &$ 229$&4 &$   2$ & $-284$ & $  78$ & $101$ & $ -64$ &$-101$ & 1.3 & 8.6 & 0.74 & 1.3  \nl
 57939 &$ -98$&2 &$-284$ & $-144$ & $  -7$ & $295$ & $  76$ &$  77$ & 1.2 & 19.2 & 0.88 & 0.1  \nl
 58229 &$ 168$&1 &$  68$ & $-194$ & $  57$ & $ 92$ & $  26$ &$  63$ & 0.4 &  8.9 & 0.91 & 0.6  \nl
 58357 &$  47$&2 &$ 150$ & $-163$ & $  44$ & $167$ & $  57$ &$  72$ & 1.0 & 10.8 & 0.83 & 0.5  \nl
 59109 &$  58$&2 &$  94$ & $-255$ & $ -85$ & $132$ & $ -35$ &$ -92$ & 0.6 & 9.3 & 0.89 & 1.1  \nl
 59239 &$  19$&1 &$   7$ & $-101$ & $  76$ & $141$ & $ 119$ &$ 141$ & 3.1 & 8.7 & 0.47 & 1.0  \nl
 59330 &$ -37$&3 &$ -45$ & $   4$ & $ -18$ & $229$ & $ 224$ &$ 225$ & 7.4 & 9.7 & 0.13 & 0.2 \nl
 59750 &$   5$&2 &$ -61$ & $ -58$ & $ -53$ & $181$ & $ 162$ &$ 170$ & 4.6 & 9.1 & 0.33 & 0.5  \nl
 60551 &$  -3$&2 &$  85$ & $ -30$ & $  -9$ & $208$ & $ 190$ &$ 190$ & 5.5 & 10.0 & 0.30 & 0.1  \nl
 60632 &$ 155$&2 &$-117$ & $-215$ & $  56$ & $130$ & $	5$ &$  56$ & 0.1 & 9.7 & 0.99 & 0.9  \nl
 60719 &$   6$&1 &$ 128$ & $ -98$ & $ -14$ & $178$ & $ 122$ &$ 123$ & 2.8 & 10.4 & 0.57 & 0.3    \nl
 61824 &$ 140$&1 &$  14$ & $-109$ & $ 120$ & $164$ & $ 111$ &$ 164$ & 3.0 & 8.9 & 0.50 & 2.6  \nl
 62747 &$  25$&1 &$ 121$ & $-510$ & $-275$ & $418$ & $-290$ &$-400$ & 7.1 & 32.9 & 0.64 & 18.0  \nl
 62882 &$ 152$&9 &$-278$ & $-194$ & $-103$ & $298$ & $  26$ &$ 106$ & 0.3 & 19.1 & 0.96 & 2.5 \nl
 63970 &$ -39$&3 &$ -29$ & $   8$ & $ -32$ & $232$ & $ 228$ &$ 230$ & 7.8 & 9.3 & 0.09 & 0.3 \nl
 64115 &$  74$&5 &$ 180$ & $ -88$ & $ 129$ & $259$ & $ 132$ &$ 185$ & 3.1 & 13.3 & 0.63 & 2.6  \nl
 64426 &$  49$&2 &$  73$ & $ -56$ & $  64$ & $190$ & $ 164$ &$ 176$ & 4.6 & 9.4 & 0.34 & 0.7 \nl
 65268 &$ -28$&1 &$  28$ & $ -10$ & $ -16$ & $213$ & $ 210$ &$ 211$ & 7.1 & 8.9 & 0.11 & 0.1 \nl
 66246 &$-101$&8 &$ 112$ & $ -65$ & $ -98$ & $215$ & $ 155$ &$ 183$ & 4.3 & 9.7 & 0.39 & 2.0 \nl
 66509 &$ -45$&2 &$ -71$ & $ -31$ & $ -54$ & $209$ & $ 189$ &$ 197$ & 5.6 & 9.6 & 0.26 & 0.5 \nl
 66665 &$ -25$&2 &$-158$ & $ -85$ & $ -76$ & $221$ & $ 135$ &$ 155$ & 3.1 & 11.8 & 0.59 & 1.0 \nl
 66815 &$ -88$&3 &$  65$ & $  -6$ & $ -65$ & $232$ & $ 214$ &$ 223$ & 6.7 & 10.0 & 0.20 & 0.7 \nl
 68594 &$ -25$&1 &$ 124$ & $-205$ & $  22$ & $126$ & $  15$ &$  27$ & 0.2 & 9.7 & 0.96 & 0.3 \nl
 68796 &$   7$&1 &$  26$ & $   5$ & $   7$ & $226$ & $ 225$ &$ 225$ & 7.7 & 9.1 & 0.08 & 0.1 \nl
 68807 &$ 155$&7 &$ -17$ & $-193$ & $ 111$ & $116$ & $  27$ &$ 115$ & 0.4 & 8.5 & 0.90 & 1.6 \nl
 69746 &$-121$&8 &$ 290$ & $-185$ & $  14$ & $292$ & $  35$ &$  38$ & 0.5 & 18.5 & 0.94 & 1.9 \nl
 70647 &$ 149$&1 &$-121$ & $-227$ & $ -39$ & $128$ & $  -7$ &$ -40$ & 0.2 & 9.2 & 0.97 & 0.8 \nl
 70681 &$ -47$&1 &$  14$ & $ -34$ & $ -69$ & $199$ & $ 186$ &$ 198$ & 6.0 & 8.5 & 0.17 & 0.7 \nl
 71886 &$ -16$&3 &$   3$ & $  -1$ & $  -4$ & $219$ & $ 219$ &$ 219$ & 7.8 & 8.5 & 0.04 & 0.1 \nl
 71887 &$  -6$&3 &$  -2$ & $  30$ & $  -4$ & $250$ & $ 250$ &$ 250$ & 8.5 & 10.2 & 0.09 & 0.1 \nl
 71939 &$ -12$&3 &$  -4$ & $  34$ & $  -3$ & $254$ & $ 254$ &$ 254$ & 8.5 & 10.5 & 0.11 & 0.1 \nl
 72461 &$  33$&2 &$-131$ & $ -94$ & $  26$ & $184$ & $ 126$ &$ 129$ & 2.9 & 10.5 & 0.57 & 0.3 \nl
 73385 &$ 177$&2 &$ -51$ & $-354$ & $  62$ & $156$ & $-134$ &$-147$ & 3.5 & 8.8 & 0.43 & 0.6 \nl
 73960 &$-278$&1 &$  -6$ & $ -61$ & $-282$ & $324$ & $ 159$ &$ 324$ & 8.0 & 11.5 & 0.18 & 8.6 \nl
\pagebreak
 74033 &$ -61$&2 &$ 105$ & $-112$ & $  31$ & $154$ & $ 108$ &$ 113$ & 2.4 & 9.7 & 0.60 & 0.3 \nl
 74067 &$ -59$&1 &$  20$ & $ -53$ & $ -60$ & $178$ & $ 167$ &$ 177$ & 5.0 & 8.6 & 0.26 & 0.6 \nl
 74079 &$  19$&3 &$ -24$ & $  10$ & $  20$ & $232$ & $ 230$ &$ 231$ & 8.0 & 9.2 & 0.07 & 0.2 \nl
 74234 &$ 312$&2 &$-312$ & $-495$ & $ -58$ & $420$ & $-275$ &$-281$ & 4.9 & 47.9 & 0.81 & 1.7 \nl
 74235 &$ 311$&2 &$-312$ & $-495$ & $ -59$ & $420$ & $-275$ &$-281$ & 4.9 & 47.7 & 0.81 & 1.7 \nl
 76976 &$-171$&2 &$ 240$ & $-239$ & $  48$ & $246$ & $ -19$ &$ -52$ & 0.2 & 14.7 & 0.97 & 0.6 \nl
 77946 &$ -65$&1 &$ 131$ & $-118$ & $ 101$ & $194$ & $ 102$ &$ 143$ & 2.2 & 10.4 & 0.65 & 1.5 \nl
 78640 &$-153$&2 &$-117$ & $-254$ & $ -18$ & $123$ & $ -34$ &$ -38$ & 0.5 & 9.7 & 0.90 & 0.2 \nl
 80837 &$ -48$&2 &$ -94$ & $-251$ & $ -71$ & $122$ & $ -31$ &$ -78$ & 0.5 & 9.3 & 0.90 & 0.8 \nl
 81170 &$-170$&2 &$  79$ & $-159$ & $-128$ & $163$ & $  61$ &$ 142$ & 1.2 & 9.4 & 0.77 & 2.2 \nl
 81461 &$ -39$&2 &$  -3$ & $-103$ & $ -12$ & $118$ & $ 117$ &$ 118$ & 2.9 & 8.4 & 0.49 & 0.1 \nl
 85007 &$  33$&5 &$ -47$ & $  26$ & $  10$ & $251$ & $ 246$ &$ 247$ & 7.9 & 10.8 & 0.16 & 0.1 \nl
 85378 &$ -73$&2 &$  56$ & $ -93$ & $  72$ & $156$ & $ 127$ &$ 146$ & 3.2 & 8.9 & 0.47 & 0.8 \nl
 85757 &$   4$&2 &$ -24$ & $ -69$ & $  72$ & $169$ & $ 151$ &$ 167$ & 4.3 & 8.5 & 0.33 & 0.7 \nl
 85855 &$-285$&2 &$ 104$ & $-424$ & $ 155$ & $277$ & $-204$ &$-257$ & 6.1 & 12.0 & 0.32 & 3.2 \nl
 86013 &$-143$&1 &$ 160$ & $ -53$ & $  83$ & $246$ & $ 167$ &$ 186$ & 4.0 & 12.7 & 0.52 & 1.2 \nl
 86431 &$  34$&2 &$-212$ & $-104$ & $  90$ & $258$ & $ 116$ &$ 147$ & 2.4 & 14.4 & 0.72 & 1.5 \nl
 86443 &$-398$&2 &$ 357$ & $-238$ & $  84$ & $368$ & $ -18$ &$ -86$ & 0.3 & 29.9 & 0.98 & 2.1 \nl
 87693 &$-243$&2 &$ -50$ & $-372$ & $ -32$ & $163$ & $-152$ &$-155$ & 4.1 & 8.8 & 0.36 & 0.3 \nl
 88010 &$ 186$&2 &$-277$ & $-272$ & $  39$ & $284$ & $ -52$ &$ -65$ & 0.8 & 17.8 & 0.92 & 0.5 \nl
 88039 &$ -16$&1 &$  -6$ & $-100$ & $  83$ & $146$ & $ 120$ &$ 146$ & 3.1 & 8.5 & 0.46 & 0.9 \nl
 91058 &$ -36$&5 &$  57$ & $  17$ & $ -31$ & $246$ & $ 237$ &$ 239$ & 7.4 & 10.7 & 0.18 & 0.3 \nl
 92167 &$-181$&1 &$  61$ & $ -68$ & $-284$ & $328$ & $ 152$ &$ 322$ & 6.8 & 13.0 & 0.32 & 9.4 \nl
 92532 &$ -13$&2 &$  13$ & $  20$ & $ -35$ & $243$ & $ 240$ &$ 242$ & 8.4 & 9.5 & 0.06 & 0.3 \nl
 92781 &$  20$&5 &$-111$ & $-133$ & $ -25$ & $143$ & $  87$ &$  90$ & 1.8 & 9.6 & 0.69 & 0.2 \nl
 94449 &$ -66$&2 &$-153$ & $-301$ & $  60$ & $183$ & $ -81$ &$-101$ & 1.5 & 10.8 & 0.75 & 0.7 \nl
 96115 &$-124$&1 &$ -40$ & $-143$ & $ -57$ & $104$ & $  77$ &$  96$ & 1.6 & 8.6 & 0.68 & 0.5 \nl
 96185 &$-167$&2 &$  56$ & $-150$ & $  63$ & $110$ & $  70$ &$  94$ & 1.4 & 8.8 & 0.72 & 0.6 \nl
 97023 &$  -9$&5 &$ -40$ & $ -16$ & $ -38$ & $212$ & $ 204$ &$ 208$ & 6.6 & 9.0 & 0.15 & 0.3 \nl
 97468 &$-181$&4 &$ 137$ & $-196$ & $-106$ & $175$ & $  24$ &$ 108$ & 0.3 & 9.8 & 0.95 & 1.5 \nl
 98020 &$-193$&2 &$ 143$ & $-102$ & $  69$ & $198$ & $ 118$ &$ 136$ & 2.6 & 10.9 & 0.61 & 0.8 \nl
 98532 &$ -15$&1 &$ -82$ & $-115$ & $  51$ & $142$ & $ 105$ &$ 116$ & 2.4 & 9.2 & 0.58 & 0.5 \nl
 99423 &$-115$&3 &$  21$ & $ -92$ & $ 107$ & $168$ & $ 128$ &$ 167$ & 3.5 & 8.6 & 0.43 & 1.4 \nl
 99938 &$-111$&9 &$ 106$ & $ -64$ & $ -35$ & $192$ & $ 156$ &$ 160$ & 4.0 & 10.1 & 0.43 & 0.3 \nl
100568 &$-172$&2 &$ 147$ & $-231$ & $ -63$ & $160$ & $ -11$ &$ -64$ & 0.2 & 10.5 & 0.97 & 0.8 \nl
100792 &$-248$&2 &$  68$ & $-264$ & $ -22$ & $ 84$ & $ -44$ &$ -49$ & 0.8 & 8.9 & 0.84 & 0.2 \nl
101346 &$ -68$&7 &$  63$ & $ -26$ & $   3$ & $204$ & $ 194$ &$ 195$ & 5.8 & 9.4 & 0.24 & 0.0 \nl
101382 &$  -6$&2 &$  20$ & $  10$ & $  48$ & $236$ & $ 230$ &$ 235$ & 8.1 & 9.2 & 0.06 & 0.4 \nl
103269 &$-131$&2 &$ -80$ & $-131$ & $ -87$ & $148$ & $  89$ &$ 124$ & 2.0 & 9.2 & 0.64 & 1.1 \nl
104659 &$ -44$&2 &$-100$ & $-102$ & $ -51$ & $163$ & $ 118$ &$ 129$ & 2.8 & 9.6 & 0.55 & 0.5 \nl
104660 &$-102$&1 &$ 119$ & $ -69$ & $ -30$ & $194$ & $ 151$ &$ 154$ & 3.7 & 10.5 & 0.48 & 0.3 \nl
105888 &$ -85$&2 &$  26$ & $-113$ & $ -36$ & $116$ & $ 107$ &$ 113$ & 2.5 & 8.5 & 0.54 & 0.3 \nl
106947 &$-104$&5 &$  55$ & $ -59$ & $  75$ & $186$ & $ 161$ &$ 177$ & 4.6 & 9.0 & 0.33 & 0.8 \nl
107975 &$  19$&1 &$ -22$ & $  28$ & $  -0$ & $249$ & $ 248$ &$ 248$ & 8.3 & 10.2 & 0.10 & 0.0 \nl
109067 &$-201$&2 &$  27$ & $-208$ & $  39$ & $ 48$ & $  12$ &$  40$ & 0.2 & 8.6 & 0.96 & 0.3 \nl
109390 &$ -16$&2 &$-122$ & $-121$ & $  17$ & $158$ & $  99$ &$ 100$ & 2.2 & 9.8 & 0.64 & 0.4 \nl
109558 &$-291$&2 &$ 293$ & $-265$ & $  12$ & $296$ & $ -45$ &$ -47$ & 0.7 & 19.4 & 0.93 & 0.2 \nl
112796 &$-116$&4 &$-277$ & $-234$ & $  82$ & $289$ & $ -14$ &$ -83$ & 0.1 & 18.0 & 0.99 & 1.6 \nl
114271 &$ -32$&10&$ -12$ & $ -92$ & $   1$ & $128$ & $ 128$ &$ 128$ & 3.3 &  8.5 & 0.44 & 0.1 \nl
114962 &$  16$&2 &$-392$ & $-317$ & $ -53$ & $407$ & $ -97$ &$-111$ & 1.4 & 43.2 & 0.94 & 1.5 \nl
115167 &$-246$&5 &$ 291$ & $-439$ & $ -55$ & $369$ & $-219$ &$-226$ & 4.3 & 30.1 & 0.75 & 1.3 \nl
115610 &$ -42$&3 &$-123$ & $ -44$ & $ -65$ & $224$ & $ 176$ &$ 188$ & 4.6 & 11.2 & 0.41 & 0.8 \nl
115949 &$-119$&1 &$-152$ & $-133$ & $ -59$ & $185$ & $  87$ &$ 105$ & 2.1 & 11.0 & 0.69 & 0.9 \nl
116082 &$  27$&3 &$  34$ & $  20$ & $ -22$ & $244$ & $ 240$ &$ 241$ & 8.0 & 10.2 & 0.12 & 0.2 \nl
117029 &$ -71$&1 &$  62$ & $ -92$ & $  89$ & $168$ & $ 128$ &$ 156$ & 3.3 &  9.1 & 0.46 & 1.1 \nl
117041 &$ -86$&2 &$ 239$ & $-166$ & $-190$ & $310$ & $  54$ &$ 198$ & 1.2 & 52.0 & 0.95 & 39.3 \nl
G97-40 &$   2$&2 &$-140$ & $-588$ & $  85$ & $403$ & $-368$ &$-378$ & 8.2 & 40.4 & 0.7 & 2.8 \nl
G110-43&$  26$&2 &$ -64$ & $-227$ & $-160$ & $172$ & $  -7$ &$-160$ & 0.1 & 13.2 & 1.0 & 4.6 \nl
G88-42 &$ 390$&2 &$ 138$ & $-643$ & $ 144$ & $468$ & $-423$ &$-447$ & 8.4 & 76.3 & 0.8 & 8.9 \nl
G90-36 &$ 268$&2 &$ 206$ & $-511$ & $ -40$ & $359$ & $-291$ &$-294$ & 6.1 & 29.8 & 0.7 & 0.8 \nl
G114-42&$ -87$&2 &$-358$ & $-165$ & $ -70$ & $369$ & $  55$ &$  89$ & 0.6 & 31.8 & 1.0 & 1.7 \nl
G116-53&$ 114$&2 &$-337$ & $ -91$ & $ 107$ & $376$ & $ 129$ &$ 168$ & 3.4 & 42.6 & 0.9 & 5.4 \nl
G197-30&$ 137$&2 &$ 413$ & $-163$ & $  44$ & $419$ & $  57$ &$  72$ & 0.7 & 49.5 & 1.0 & 1.2 \nl
G166-37&$ 369$&2 &$-295$ & $-126$ & $ 341$ & $461$ & $  94$ &$ 354$ & 2.2 & 53.6 & 0.9 & 39.5 \nl
G15-13 &$ 219$&2 &$-223$ & $-428$ & $ 163$ & $346$ & $-208$ &$-264$ & 4.8 & 22.0 & 0.6 & 5.3 \nl
G16-25 &$  23$&2 &$ -73$ & $-557$ & $ 160$ & $380$ & $-337$ &$-373$ & 8.0 & 26.9 & 0.5 & 5.7 \nl
G93-1  &$-121$&2 &$-332$ & $-462$ & $ -91$ & $421$ & $-242$ &$-259$ & 3.8 & 46.9 & 0.8 & 4.4 \nl
\enddata
\tablenotetext{a}{The first 9 columns of data for the Stephens 1999 stars (those with Giclas numbers) are taken directly from the Carney \etal 1994 survey. The orbital parameters ($R_{min}$, $R_{max}$, $e$, and $|Z_{max}|$ were recalculated using the same galactic potential as the F00 stars.}
\tablenotetext{b}{References:  (1) HIC, (2) CLLA, (3) F00 spectra, (4) Bond 1980, 
(5) Sandage \& Fouts 1987, (6) Beers \etal 2000, (7) Norris 1986, (8) Chiba 
\& Yoshii 1998, (9) Beers \& Sommer-Larson 1995, (10) Stetson 1983 and 
(11) Bartkevicius \etal 1992.}
\end{deluxetable}

%%%%%%%%%%%%
% Table 2  %
%%%%%%%%%%%%

\clearpage
\begin{deluxetable}{lrrrrrrrrr}
\tablenum{2}
\tablewidth{0pt}
\tablecaption{Abundance Ratios as a Function of Rest-Frame Velocity for $-1 >$ [Fe/H] $> -2$}
\tablehead{  &\multicolumn{3}{c}{$v_{RF} < 150$ } & \multicolumn{3}{c}{$150 < v_{RF} < 300$ } & \multicolumn{3}{c}{$v_{RF} > 300$ } \nl
\colhead{Quantity} & \colhead{Mean} & \colhead{sdom} & \colhead{n} & \multicolumn{1}{c}{Mean} & \colhead{sdom} & \colhead{n} & \multicolumn{1}{c}{Mean} & \colhead{sdom} & \colhead{n} }
\startdata
$T_{eff}$ (K) & 5354 & 138 & 21 &  5218   &  85  & 31 &  5300   &  98  & 21 \nl
log(g)    &    3.1  &  0.3 & 21 &     3.2 &  0.2 & 31 &     4.0 &  0.2 & 21 \nl
[Fe/H]    & $-1.61$ & 0.05 & 21 & $-1.49$ & 0.06 & 31 & $-1.52$ & 0.05 & 21 \nl
log n(Li) & $+2.07$ & 0.13 & 12 & $+1.55$ & 0.16 & 17 & $+1.48$ & 0.28 & 17 \nl
[Na/Fe]   & $-0.07$ & 0.03 & 21 & $-0.10$ & 0.03 & 30 & $-0.34$ & 0.05 & 21 \nl
[Mg/Fe]   & $+0.44$ & 0.02 & 21 & $+0.41$ & 0.02 & 31 & $+0.22$ & 0.03 & 21 \nl
[Al/Fe]   & $+0.24$ & 0.05 &  7 & $+0.24$ & 0.04 & 14 & $-0.02$ & 0.13 &  5 \nl
[Si/Fe]   & $+0.47$ & 0.02 & 20 & $+0.40$ & 0.02 & 30 & $+0.31$ & 0.03 & 17 \nl
[Ca/Fe]   & $+0.34$ & 0.02 & 21 & $+0.32$ & 0.01 & 31 & $+0.24$ & 0.02 & 21 \nl
[Ti/Fe]   & $+0.31$ & 0.02 & 20 & $+0.29$ & 0.02 & 31 & $+0.26$ & 0.03 & 21 \nl
[V/Fe]    & $-0.01$ & 0.05 &  7 & $+0.00$ & 0.02 & 27 & $-0.02$ & 0.10 &  8 \nl
[Cr/Fe]   & $-0.08$ & 0.02 & 20 & $-0.04$ & 0.01 & 30 & $-0.02$ & 0.01 & 21 \nl
[Ni/Fe]   & $+0.01$ & 0.01 & 21 & $+0.00$ & 0.01 & 30 & $-0.06$ & 0.02 & 21 \nl
[Y/Fe]    & $-0.02$ & 0.04 & 16 & $-0.05$ & 0.04 & 24 & $-0.17$ & 0.04 & 19 \nl
[Zr/Fe]   & $+0.25$ & 0.08 & 13 & $+0.26$ & 0.05 & 21 & $+0.19$ & 0.05 &  6 \nl
[Ba/Fe]   & $+0.08$ & 0.04 & 21 & $+0.07$ & 0.03 & 30 & $-0.02$ & 0.03 & 21 \nl
[Eu/Fe]   & $+0.36$ & 0.05 &  7 & $+0.41$ & 0.05 & 15 & $+0.44$ & 0.06 &  6 \nl
[Y/Zr]    & $-0.29$ & 0.06 & 13 & $-0.32$ & 0.03 & 21 & $-0.35$ & 0.10 &  6 \nl
[Ba/Eu]   & $-0.21$ & 0.11 &  7 & $-0.40$ & 0.06 & 15 & $-0.49$ & 0.07 &  6 \nl
$v_{RF}$ (km $s^{-1}$) & 120 & 5  & 21 &     212 &    9 & 31 &     389 &    8 & 21 \nl
$v_{ROT}$ (km $s^{-1}$)&  47 & 18 & 21 &      39 &   20 & 31 &  $-108$ &   43 & 21 \nl
$h$ (km $s^{-1}$)&      43 &   20 & 21 &      42 &   24 & 31 &   $-88$ &   57 & 21 \nl
$R_{max}$ (kpc)& 9.1 & 0.1 & 21 &    14.3 &  2.4 & 31 &    37.4 &  3.1 & 21 \nl
$|Z_{max}|$ (kpc)& 0.6 & 0.1 & 21 &   3.1 &  2.2 & 31 &     6.2 &  1.9 & 21 \nl
$e$       &    0.72 & 0.04 & 21 &    0.66 & 0.05 & 31 &    0.76 & 0.05 & 21 \nl
\enddata
\end{deluxetable}

\end{document}